\documentclass[usenatbib]{aa}

\usepackage{txfonts}
\usepackage{graphicx}
\usepackage[update,prepend]{epstopdf}
\graphicspath{images}
\usepackage{amsmath}
\usepackage{caption}
\usepackage{subcaption}
\usepackage{adjustbox}
\usepackage{ragged2e}
\usepackage{gensymb}
\usepackage{threeparttable}

\begin{document}

\title{Analysis of the public HARPS/ESO spectroscopic archive}
\subtitle{Jupiter-like planets around HD\,103891 and HD\,105779}

\author{
 K.R. Sreenivas\inst{1}
 \and V. Perdelwitz\inst{1}
 \and L. Tal-Or\inst{1,2}
 \and T. Trifonov\inst{3}
 \and S. Zucker\inst{4} 
 \and T. Mazeh\inst{5} 
}
\institute{
 Department of Physics, Ariel University, Ariel 40700, Israel\\
 \email{levtalor@ariel.ac.il}
 \and Astrophysics Geophysics And Space Science Research Center, Ariel University, Ariel 40700, Israel
 \and Max-Planck-Institut f\"ur Astronomie, K\"onigstuhl 17, D-69117 Heidelberg, Germany
 \and The Porter School of Environment and Earth Sciences, Raymond and Beverly Sackler Faculty of Exact Sciences, Tel Aviv University, Tel Aviv 6997801, Israel 
 \and School of Physics and Astronomy, Raymond and Beverly Sackler Faculty of Exact Sciences, Tel Aviv University, Tel Aviv, 6997801, Israel
}

\date{Received 8 November 2021 / Revised 16 December 2021 / Accepted xxx}
 
\abstract
 {
 }
 {We use the recently published database \citep{Trifonov2020} of radial velocities (RVs) that were derived from fifteen years of HARPS/ESO observations to search for planet candidates.
 }
 {For targets with sufficient RV data, we apply an automated algorithm to identify significant periodic signals and fit a Keplerian model for orbital estimates. We also search the auxiliary data of stellar-activity indices and compare our findings with existing literature, to detect periodic RV signals that have no counterpart in the activity timeseries. The most convincing signals are then manually inspected to designate additional false planet detection, focusing the search on long-period ($P>1\,000$\,d) massive candidates around FGK dwarf stars.
 }
 {We identify two Jupiter analogs, in orbit around the slightly evolved F8V star HD\,103891 and the Solar-like star HD\,105779. We use nested sampling to derive their orbital parameters, and find their orbital periods to be $1919\pm16$\,d and $2412\pm54$\,d, while their minimum masses are $1.44\pm0.02$\,M$_{\rm Jup}$ and $0.64\pm0.06$\,M$_{\rm Jup}$, respectively. While the orbit of HD\,103891\,b is slightly eccentric ($e=0.31\pm0.03$), that of HD\,105779\,b is likely circular ($e<0.16$).
 }
 {With minimum astrometric signatures of $\sim 59$ and $\sim 42$\,$\mu$as, HD\,103891\,b and HD\,105779\,b join the growing sample of planets whose exact masses may soon be derived with Gaia astrometry. This finding also highlights the importance of long-term RV surveys to study planetary occurrence beyond the snow line of Solar-like stars.
 }

\keywords{Techniques: radial velocities -- Astronomical data bases -- planetary systems}

\authorrunning{Sreenivasan et al.}
\titlerunning{Jupiter-like planets around HD\,103891 and HD\,105779}

\maketitle

\section{Introduction}
\label{sec1}

Almost $4\,900$ exoplanets have been discovered in the last thirty years\footnote{\label{note1}\url{http://exoplanet.eu/catalog/}, as of 16.12.2021}, with ${\sim}70\%$ of them being discovered by transit method, and another ${\sim}20$\% discovered via the radial velocity (RV) technique. As opposed to transiting-planet surveys such as the {\it Kepler} space telescope \citep{Borucki2008} and {\it Transiting Exoplanet Survey Satellite} missions \citep[TESS;][]{Ricker2015} with the capability of monitoring $>10^5$ stars simultaneously, high-resolution Doppler spectroscopy can only be carried out one star at a time, limiting the number of stars observed per night to low double digits. However, while transit searches have extremely low probability of finding long-period planets ($P>1\,000$\,d), legacy RV surveys, which monitor a sample of stars over decades, enable the discovery of cold, massive planets with a semi-major axis of up to $\sim10$\,au \citep[e.g.,][]{Hatzes2016}. Together with Gaia astrometry \citep[][]{Gaia2016,Lindegren2018,lindegren2021} and next-generation direct-imaging techniques \citep[e.g.,][]{starck2020, li2021}, the RV method may soon enable an almost complete mapping of massive planets beyond the snow line of nearby stars, regardless of orbital orientation \citep[e.g.,][]{Perryman2014,Sozzetti2014,Venner2021}.

This is a prerequisite for estimating the occurrence rate of cold massive planets around FGKM dwarf stars \citep[e.g.,][]{Fernandes2019,Fulton2021,Sabotta2021arXiv}, which may promote our understanding of planetary-system formation.

HARPS \citep{mayor03} is a high-resolution visible-light echelle spectrograph installed at the ESO $3.6$-m telescope in La Silla, Chile, which led the RV planet-discovery efforts in the southern hemisphere since 2003. With its proven long-term stability and simultaneous calibration it has been demonstrated to reach and maintain a RV accuracy of $\sim1$\,m\,s$^{-1}$ over years \citep[e.g.,][]{Pepe2014}. Despite its remarkable performance, HARPS data are known to contain certain kinds of systematic effects. For example, \citet{Dumusque2015ApJ...808..171D} have found a one-year variability caused by block stitches of its CCD detector, and a fiber upgrade in May 2015 \citep{LoCurto2015} caused a spectral-type dependent change in RV zero point of $\sim10$\,m\,s$^{-1}$.

HARPS RVs derived by the data-reduction software (DRS) via cross-correlation functions \citep[CCFs,][]{Queloz2001,Pepe2002} can be retrieved directly from the spectrum file headers, available at the ESO archive\footnote{\label{note2}\url{http://archive.eso.org/wdb/wdb/adp/phase3_main/form?}}. 
In a recent work, \citet{Trifonov2020} derived slightly more precise RVs using the SpEctrum Radial Velocity AnaLyser \citep[SERVAL,][]{Zechmeister2018}, corrected them for HARPS' nightly zero-point variations \citep[NZPs,][]{Talor2019hires}, and made the data available in the form of the {\sc HARPS-RVBank}\footnote{\label{note3}\url{http://www.mpia.de/homes/trifonov/HARPS_RVBank.html}} (also available at the {\sc cds}\footnote{\label{note4}\url{https://cdsarc.cds.unistra.fr/viz-bin/cat/J/A+A/636/A74}}).

We make use of this large database of radial velocities, in order to search for yet-undetected planet candidates. We identify two Jovian mass planet candidates orbiting the stars HD\,103891 and HD\,105779. To validate the planetary nature of the periodic RV variability in the data, we investigate the auxiliary sets of spectroscopic activity indices as well as photometric observations. 
Section~\ref{sec2} describes the underlying data, Section~\ref{sec3} outlines the methods employed to detect and validate the planetary candidates HD\,103891\,b and HD\,105779\,b, Section~\ref{sec4} describes the results of our analysis, Section~\ref{sec5} discusses the characteristics of the host stars and planet candidates, and Section~\ref{sec6} summarizes our findings.
\section{Data}
\label{sec2}
In the following section we describe the publicly available RV and photometric data used in our analysis.

\subsection{HARPS radial velocities}

Out of the data products provided by \citet{Trifonov2020} in the {\sc HARPS-RVBank}, we selected the NZP-corrected SERVAL RVs \citep{Zechmeister2018} as our main RV library.
NZP-corrected SERVAL RVs were shown to be more precise than HARPS DRS RVs by $\sim5\%$ on average. Although for some fraction of the stars HARPS-DRS RVs could be superior to NZP-corrected SERVAL RVs \citep{yarara2021,Trifonov2020}, we favored the more precise set of RVs for our global search of planet candidates.

\subsection{Activity indicators}
Aside from sub-stellar companions, apparent RV modulations can be induced by several different phenomena, the most common of which in main sequence stars is rotational modulation due to magnetic activity. In order to verify that a signal is in fact caused by orbital motion, one needs to exclude this effect as an underlying cause of the RV modulation.
One way of doing this is to show that the spectral line profile does not change with the corresponding RV period. Since the CCF obtained during the reduction process is a proxy of spectral lines, CCF analysis is a proven method to identify stellar activity induced signals \citep{Queloz2001}. 
The HARPS pipeline provides several measures of the CCF of each spectrum, including its Full-Width at Half-Maximum (FWHM), Contrast, and Bi-sector Inverse Slope (BIS). Similar to FWHM, the Differential line width (dLW) provided by SERVAL is also a good measure of line shape variations. Importantly, SERVAL also provides the ChRomatic indeX \citep[CRX,][]{Zechmeister2018} with which one can check whether any apparent RV variability is coherent in wavelength, as a true Doppler variation should be. The CRX is basically the slope obtained by fitting a straight line to the RV-order relation of each exposure. Since echelle orders are a function of wavelength, a non-zero CRX can be indicative of a wavelength dependent phenomenon. The {\sc HARPS-RVBank} provides both DRS- and SERVAL-derived spectroscopic activity indices for each spectrum.

Chromospheric line emission in stellar spectra is also a clear sign for surface activity. The emission in a number of these lines is calculated by SERVAL and provided through the {\sc HARPS-RVBank}. Out of these, we used the H$\alpha$ and Na\,D lines, which are derived by using the equations of \citet{Kurster2003}. In addition, we used $\mathrm{R}^\prime_{HK}$, a measure of the chromospheric emission in the Ca~{\sc ii}~H\&K lines, calculated using the method introduced by \citet{Perdelwitz2021}. The latter values will soon be published in the upcoming version of the {\sc HARPS-RVBank}.

\subsection{Photometry}
Stellar activity can also manifest itself as periodic brightness variations. Active surface regions, such as starspots, modulate the brightness of a star with its period of rotation ($P_{\rm rot}$). At longer periods, Sun-like stars may have activity cycles, which are characterized by periodic variability of their spot coverage fraction. Any periodicities discovered within the photometric signals can be used to identify false-positive planet signatures in the RV data. In order to investigate the short- and long-term photometric variability (rotation periods, cycles) of the host stars of the planet candidates presented here, we used their publicly-available optical lightcurves, as listed below. 

\subsubsection{Hipparcos}
Hipparcos, launched in August 1989, was a space-based mission designed and constructed by \citep{esa1997}. During its 3 years of operation the mission obtained high quality astrometric and photometric data of $>118\,000$ stars, resulting in the Hipparcos catalog \citep{Perryman1997}. 
The \textit{Hp}-band photometry ($380$--$800$\,nm) is quite precise, with typical residuals of $5$ and $10$\,mmag for 6$^{th}$ and 8$^{th}$ magnitude star, respectively \citep{vanLeeuwen1997}. 
The Hipparcos photometry for our targets were obtained using Vizier\footnote{\label{note6}\url{https://cdsarc.cds.unistra.fr/viz-bin/cat/I/239}}. The \textit{Hp}-band photometric magnitudes for HD\,103891 and HD\,105779 contains 92 and 120 measurements, with median values 6.73 and 8.76 respectively. The data does not show any photometric variability on visual inspection.
\subsubsection{All Sky Automated Survey}
ASAS \citep{Pojmanski1997} is a ground based long term photometric survey, observing objects brighter than $V_{\rm mag}=14$. The southern station is mounted at Las Campanas observatory, Chile. Since 1997 it has been monitoring $\sim10^7$ stars in the $V$ and $I$ bands. 
Typically, each star is observed $5$--$10$ times per night, at least once in 3 days. All photometric $V$-band lightcurves of the southern hemisphere until December 2009 are publically available at the ASAS webpage\footnote{\label{note7}\url{http://www.astrouw.edu.pl/asas/?page=download}}. The ASAS V band photometry contains lightcurves produced with five different apertures. For the analysis of HD\,103891 and HD\,105779 we selected the ones with lowest scatter.

\subsubsection{Transiting Exoplanet Survey Satellite}
The Transiting Exoplanet Survey Satellite \citep{Ricker2015} is an MIT-led NASA mission capable of acquiring photometry of bright stars of large areas the sky, designed to detect exoplanet transits. It has four wide field ($24^{\rm o}$ $\times$ $24^{\rm o}$) CCDs which tile the whole sky since 2018 in a point-and-stare strategy. Each point (or sector) is continuously observed for 27 days, during which the brightness of $\sim20\,000$ stars is measured every two minutes. The data of each sector becomes publicly available through the MAST archive\footnote{\label{note8}\url{https://archive.stsci.edu/missions-and-data/tess}} immediately after data reduction. TESS observed HD\,103891 in sectors $9$ and $36$ and HD\,105779 in sectors $10$ and $36$.

\section{Data analysis}
\label{sec3}
In the following section we describe our data analysis, beginning with the candidate selection, followed by a short description of the host star properties and the periodogram analysis.

\subsection{Candidate selection}
The search for periodic RV signals in the {\sc HARPS-RVBank} was carried out by iteratively using the 
Generalized Lomb-Scargle periodogram \citep[GLS,][]{Zechmeister2009a} 
and fitting routines. Since we treated pre- and post-fiber-upgrade RVs as two separate data sets that may have different zero points (hereafter, pre and post), the GLS was modified to incorporate two offsets. 
The RV data of each star were first fitted with a null model, which only includes two offsets and two additive jitter terms \citep[e.g.,][]{Gregory2005b,Baluev2009}, and then iteratively fitted with up to eight Keplerian signals. 

For each signal, the GLS peak frequency was used as an initial guess for the next Keplerian fit, which was done by maximising the following Likelihood function:
\begin{equation}
\ln L_{\rm total} = \ln L_{\rm pre} + \ln L_{\rm post}, 
\label{eq:1}
\end{equation}
where the ln\,$L$ of the individual instruments is:
\begin{equation}
\begin{split}
 \ln L_{\rm pre/post} = -\frac 1 2 \sum_{n} \frac{(v_{n} - \mu_{\rm pre/post} - \mathrm{f}_k)^2} {S_{\rm pre/post}^2 + \sigma_n^2} \\- \frac 1 2 \sum_{n} \ln(S_{\rm pre/post}^2 + \sigma_n^2) - \frac n 2 \ln (2\pi), 
\end{split}
\label{eq:2}
\end{equation}
$n$ is the number of points in the RV dataset, $v_n$ and $\sigma_n$ are the RVs and their uncertainties, $\mu_{\rm pre/post}$ is the RV offset, $S$ is the RV jitter, and ${\rm f}_k$ is the five-parameter Keplerian model of the k-th signal. The best-fit Keplerian model was then subtracted from the data and the residuals were passed on to the next iteration. At this preliminary search stage, the RV variations of a star produced by each planet in a multi planetary system were assumed to be independent, so that upon the subtraction of each signal, the residuals represent the RV variations produced by the rest of the planets. 

Throughout this preliminary search for periodic RV signals, ln\,$L$ was used as the test statistic. After fitting the data with the null model, we added Keplerian signals either until the change in $\ln L_{\rm total}$ (hereafter, $\Delta\ln L$) became negative or until the number of free parameters exceeded the number of RVs. 

We then carried out target by target manual inspection of the most promising candidates. Among the things we checked were: the uniqueness of the peaks in their GLS periodograms, window function (i.e. possible aliases), long-term trends, signs of stellar binarity, phase coverage, and the modelling statistics, shortlisting the candidates on the basis of $\Delta\ln L$ value. For this stage we used {\sc Exo-Striker}\footnote{\label{note9}\url{https://github.com/3fon3fonov/exostriker}} fitting toolbox \citep{trifon2019}. Here, we focused our search on long-period ($P>1\,000$\,d) massive candidates around FGK dwarf stars with observational time span ($T_{\rm span}$) of at least two orbital periods and RV semi-amplitude of $>10$\,m\,s$^{-1}$.

We also searched the data for periodic RV signals that probably originate from stellar activity. Similar to the RVs, the activity indices of each star were split into pre and post data and were analysed by using GLS periodograms. A periodic RV signal was marked as activity, if significant activity-related periodicity was found to lie within $3\sigma$ of the corresponding period of the RVs.

Among the cold massive planet candidates in orbit around FGK dwarf stars we found the most promising ones to orbit HD\,103891 and HD\,105779. The physical parameters of the host stars are listed in Table \ref{tab:tab1}. In what follows, we focus on an in-detail analysis of the available data of these two stars. We defer similar analysis of additional candidates to future work.

\begin{table}[!t]
 \centering
 \setlength{\tabcolsep}{8pt} 
 \caption{Host-star parameters.}
 \label{tab:tab1}
 \begin{tabular}{l c c c c}
 \hline
 \hline
 Parameter & HD\,103891 & HD\,105779 & ref. \\ [0.1ex]
 \hline
 RA\,[deg]&179.438276&182.661171&1 \\
 DEC\,[deg]&$-$8.548616&$-$16.956497&1\\
 $\pi$\,[mas] & 18.22$\pm$0.04 & 18.15$\pm$0.03 &1\\
 $\mu_{\alpha}$\,[mas/yr] &-108.34$\pm$0.04 &-235.64$\pm$0.03 &1\\
 $\mu_{\delta}$\,[mas/yr] &21.84$\pm$0.02&-46.25$\pm$0.02&1\\
 $\epsilon_i$\,[mas] & 0.236 & 0.111 &1\\
 $\epsilon_i$\,significance & 74.1 & 9.7 &1\\
 \small ruwe & 0.98 & 1.04 &1\\
 
 G magnitude &6.45&8.50&1\\
 Spectral Type & F9 & G2 &2\\
 
 $M_{v}$& 2.79$\pm$0.20 & 4.87$\pm$0.28 &3\\
 L\,[L$_{sun}$] & 6.11$\pm$0.02 & 0.78$\pm$0.04&4\\
 V & 6.55$\pm$0.01 & 8.64$\pm$0.01 &5\\
 B - V & 0.567$\pm$0.003 & 0.614$\pm$0.001 &6\\
 
 Age\,[Gyr] &3.68$\pm$0.078&7.54$\pm$1.30&2\\
 $T_\mathrm{eff}$\,[K]&6072$\pm$20 & 5792$\pm$16&7\\
 $R_{*}$\,[R$_{\sun}$]&2.22$\pm$0.05 & 0.94$\pm$0.02 &2\\
 $M_{*}$\,[M$_{\sun}$] & 1.28$\pm$0.01 & 0.89$\pm$0.01 &2\\
 $v\sin{i}$\,[km s$^{-1}$] & 2.8$\pm$0.1 & 2.0$\pm$0.1&8\\
 $[\rm Fe/H]$\,$[\rm dex]$ &-0.19$\pm$0.01&-0.25$\pm$0.01&8\\
 log g\,[cm s$^{-2}$] &3.79$\pm$0.03& 4.36$\pm$0.03&8\\
 $\log \overline{R^\prime_{HK}}$ &-5.20&-4.94& 9\\
 P$_{\rm rot}/\sin{i}$\,[d] &40.3$\pm$2.3&23.8$\pm$1.7& 9\\
 \hline
 \end{tabular}
 \tablebib{
(1)~\citet{lindegren2021}; (2) \citet{Gomas}; (3) \citet{allende}; (4) \citet{Sousa};
(5) \citet{Perryman1997}; (6) \citet{vanLeeuwen2007b}; (7) \citet{mena}; (8) \citet{costasilva};
(9) This work.\vspace{-0.2cm}
}
\end{table}

\subsection{Periodograms, coherence analysis, and nested sampling}
\label{sec:3.3}
The two planet candidates presented here have orbital periods of $\sim2\,000$\,d, and their HARPS observations span only $\sim3$ orbital periods. In GLS periodogram analysis, sampled equidistantly in frequency space, that means low-resolution peaks (or high period uncertainty) close to the low-frequency window edge. Moreover, any long-term variability in the RV or activity-index timeseries, be it a linear trend or a periodic variability of a period longer than $T_{\rm span}$, may induce spurious low-frequency peaks in the range of the proposed orbital periods. 
To better understand the distribution of power in RV and activity index data close to the proposed planetary periods, we explored each dataset by using a $\Delta\ln L$ periodogram \citep{trifon2020b} . Here, $\Delta\ln L$ denotes the significance of the addition of a periodic term with respect to a null model composed of instrument-by-instrument offsets, jitters and linear trends. We opted for instrument-by-instrument linear trends (as opposed to a single linear trend) in order to also account for linear trends caused by non-astrophysical reasons, such as systematic instrumental variations. Since this approach is computationally heavy, we only applied it to the two most promising candidates in our list. This approach results in the suppression of power at periods larger than $2T_{\rm span}$, similar to a high-pass filter. Its advantage is that it can suppress spurious periodogram peaks while not influencing those resulting from true variability at periods smaller than $T_{\rm span}$.

One of the differences between a Keplerian signal and an activity-induced one is that, while the former is expected to be coherent over time, the latter can vary in amplitude and phase \citep{sbgls}. While the $\Delta\ln L$ of fitting a Keplerian to a coherent signal should increase upon adding RV measurements, amplitude or phase variations can lead to reduction in the $\Delta\ln L$ over time. We observed the significance evolution of the proposed planetary signals by calculating the evolution of $\Delta\ln L$ with respect to their null models.

Once we were convinced that the periodic RV variability is most probably caused by orbital motion of the host stars, we carried out Bayesian modelling of Keplerian orbits by sampling from the posterior probability distribution function (PDF) of the parameters, based on the given data. The direct determination of posterior values using Bayes theorem is difficult, due to the presence of multi-dimensional integral known as Evidence ($\mathcal{Z}$) \citep{Gregory2005b}. Usually, the PDF estimation is determined by sampling from a distribution which is proportional to it. Classical Markov-Chain Monte Carlo (MCMC) algorithms construct PDFs from the product of the likelihood and the prior functions \citep{ford2006}.
Nested sampling deals with the calculation of $\ln\mathcal{Z}$, the logarithm of normalisation factor of Bayes' theorem, which is a higher dimensional integral. It is advantageous over classical MCMC analysis in that we can use $\ln\mathcal{Z}$ to compare different models, 
and thus get the full PDF of the parameters of the most probable model \citep{skilling2004}. Since nested sampling basically focus on providing $\Delta\ln \mathcal{Z}$, we used dynamic nested sampling \citep{speagle}. The dynamic nested sampling is advantageous in that, it can be tuned to focus on posterior calculation and provide $\ln\mathcal{Z}$ as a by-product, which will allow comparison of different models. We used the python package \texttt{juliet} \citep{juliet2020} which uses Nested sampling in the modelling of the HARPS RVs of HD\,103891 and HD\,105779.

\section{Results}
\label{sec4}
In the following, we discuss the results of the two individual planet candidates derived by applying the methods described in Sect. ~\ref{sec:3.3}.

\subsection{HD\,103891}
\begin{figure}[t!]
\begin{subfigure}{0.5\textwidth}
  \includegraphics[width=0.98\linewidth]{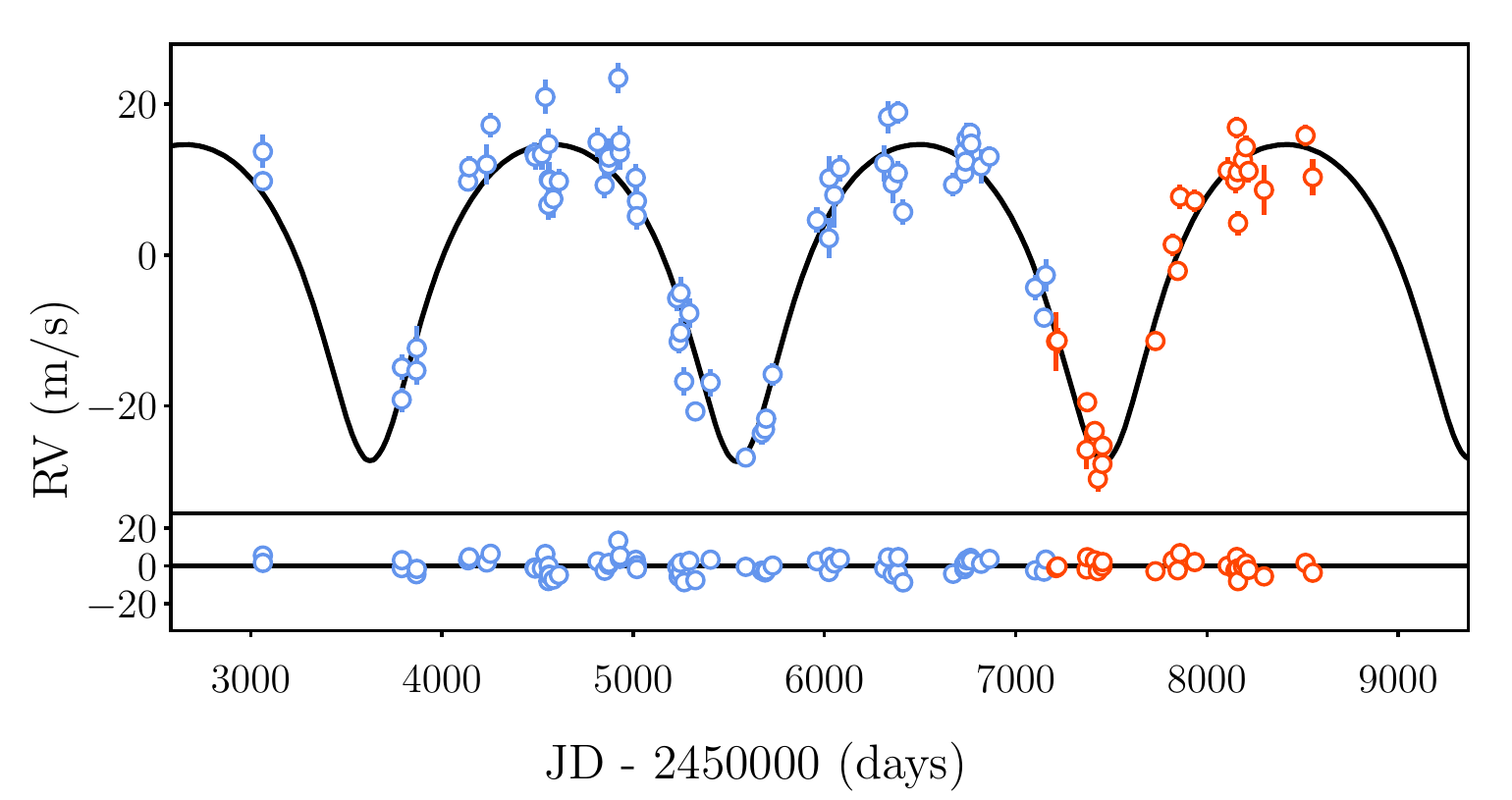}
  \label{fig:subim1}
\end{subfigure}
\hfill
\begin{subfigure}{0.5\textwidth}
  \includegraphics[width=0.98\linewidth]{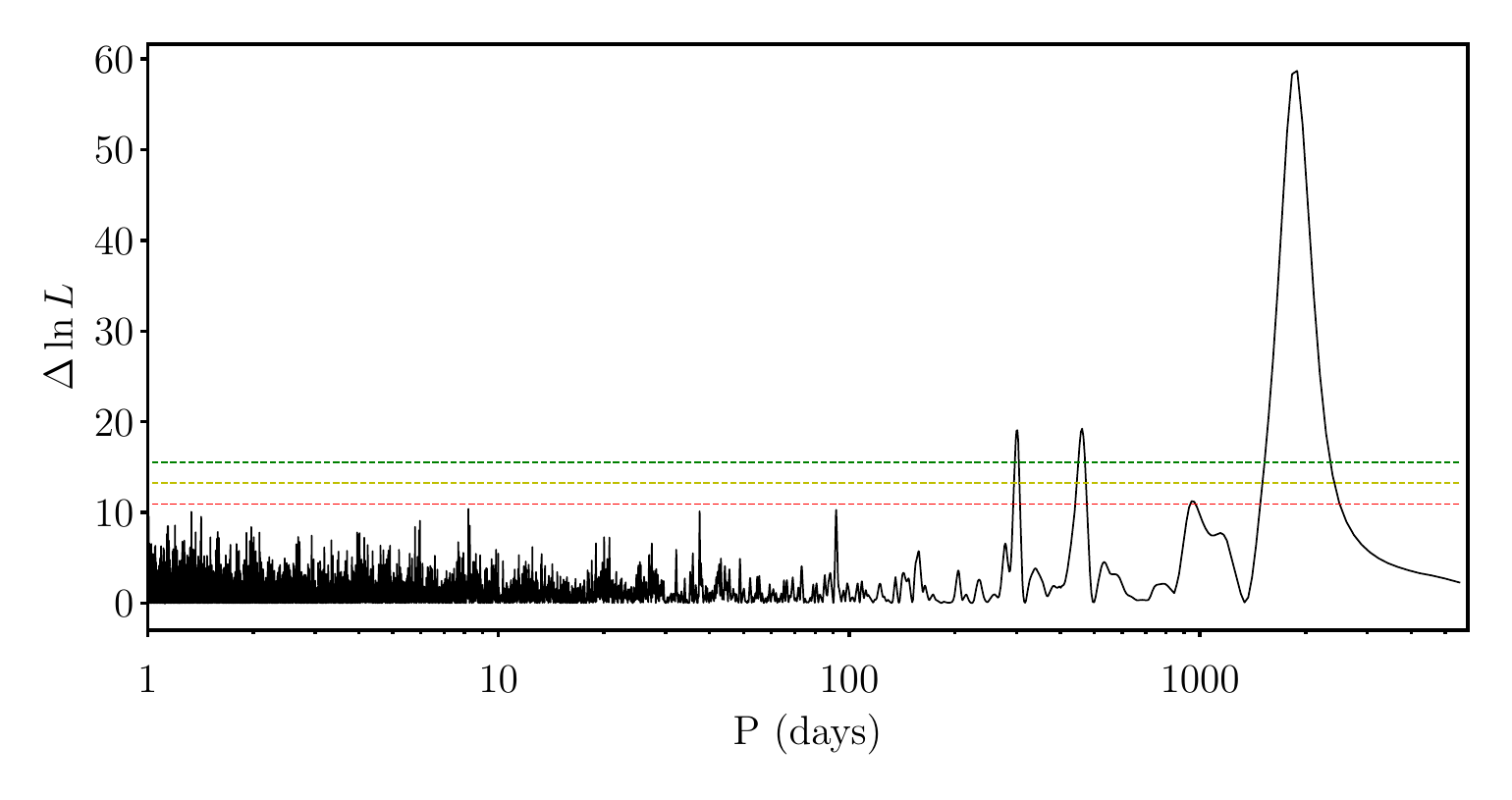}
  \label{fig:subim2}
\end{subfigure}
\hfill
\begin{subfigure}{0.5\textwidth}
  \includegraphics[width=0.98\linewidth]{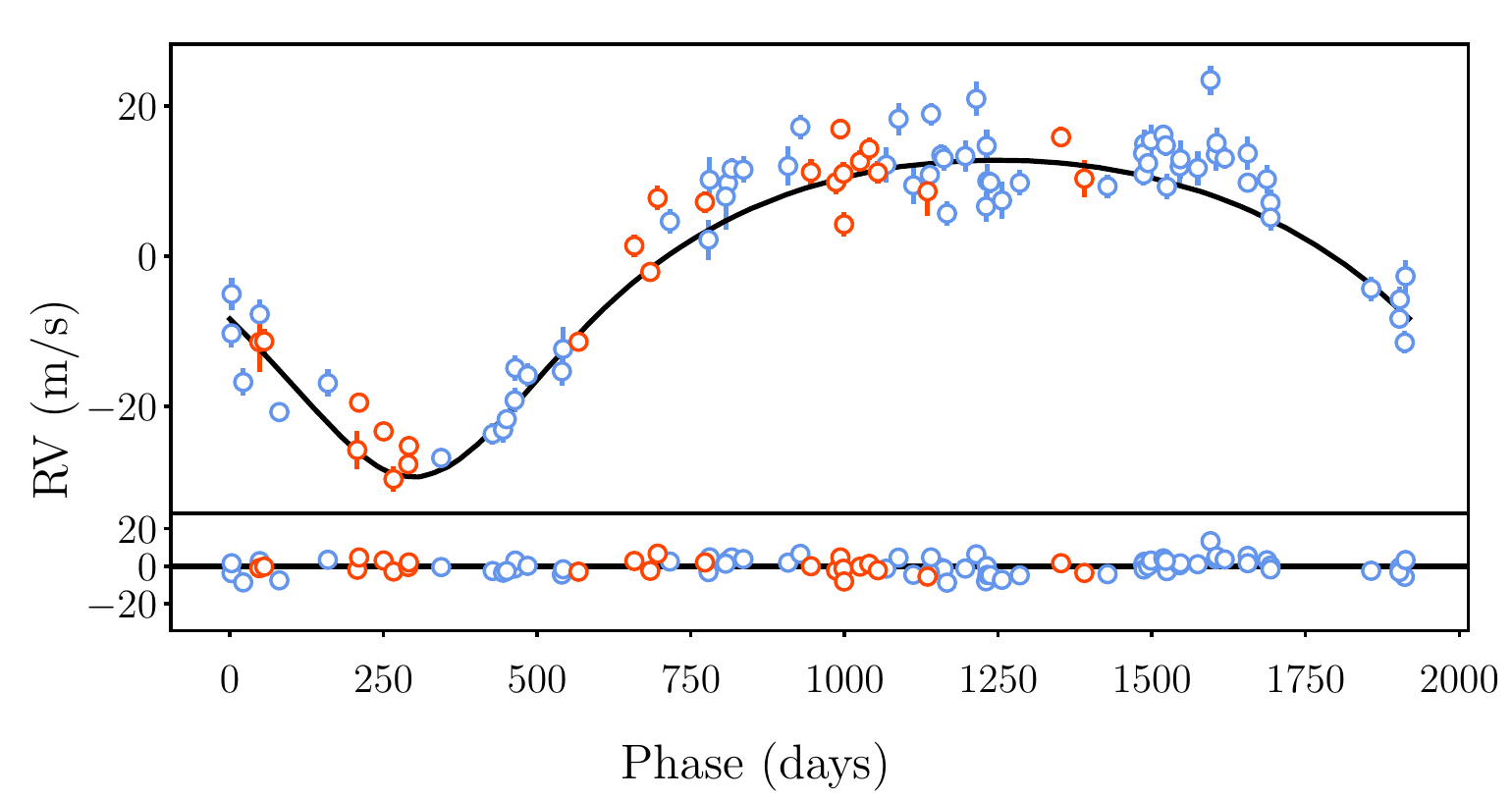}
  \label{fig:subim3}
\end{subfigure}
\hfill
\begin{subfigure}{0.5\textwidth}
  \includegraphics[width=0.98\linewidth]{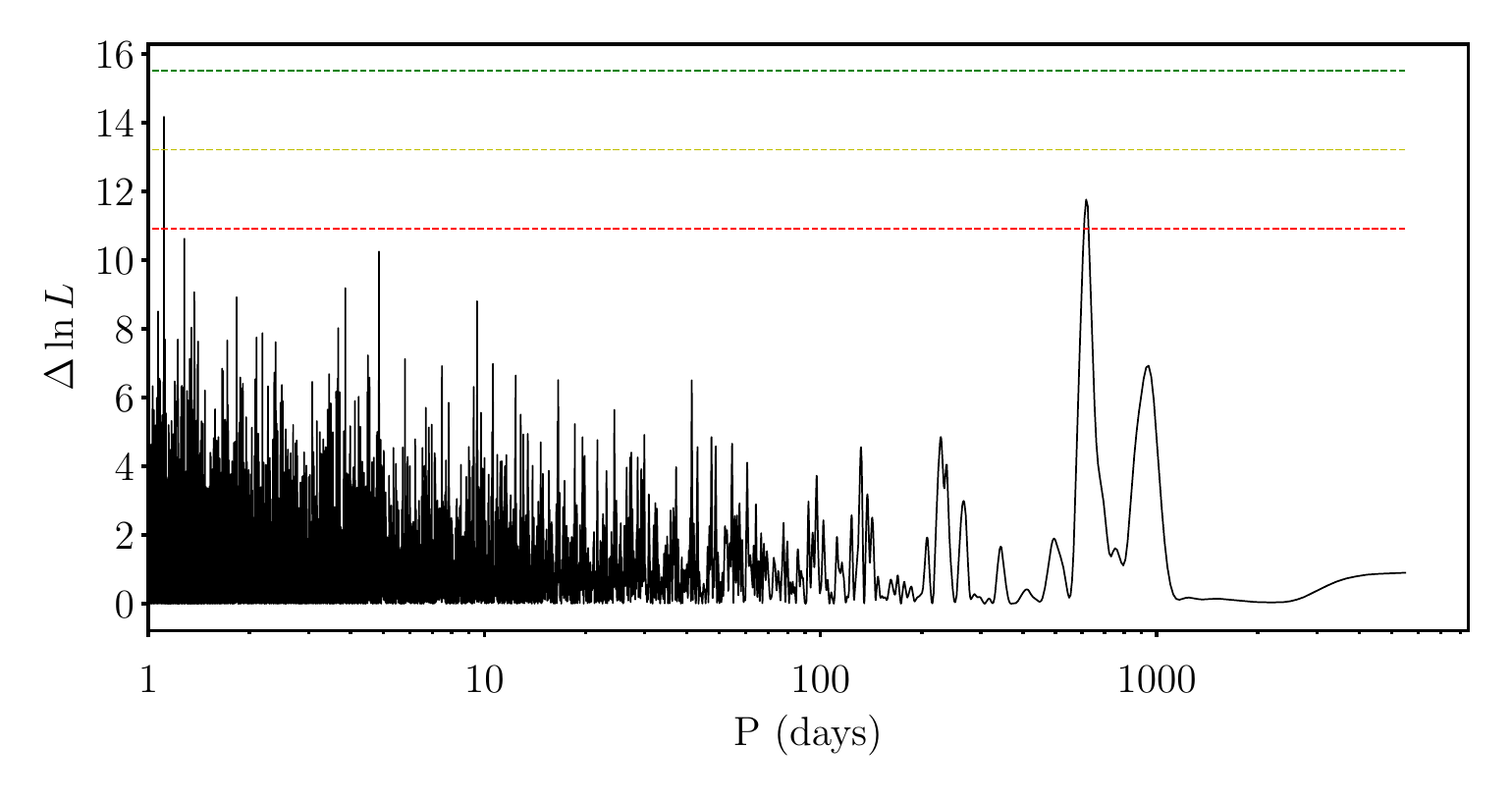}
  \label{fig:subim4}
\end{subfigure}
\hfill
 \captionsetup{width=.95\linewidth}
\caption{Radial velocities and resulting periodograms of HD\,103891. \textit{Top left:} The HARPS-SERVAL NZP-corrected RVs. pre RVs are colored blue and post RVs are colored red. The best-fit Keplerian model is shown with the solid black line. The bottom panel of this figure shows the residuals. \textit{Top right:} $\Delta\ln L$ periodogram of the input RVs. The periodogram shows a peak at $1893$\,d. The horizontal dashed lines show 0.001 (Green), 0.01 (Yellow), and 0.1 (Red) FAP values. \textit{Bottom left:} The RV data and Keplerian model, phase folded at the best-fit orbital period. \textit{Bottom right:} $\Delta\ln L$ periodogram of the residuals.}
\label{fig:1}
\end{figure}

HD\,103891 was observed with HARPS from February 2004 to April 2018. The NZP-corrected data contain 91 RV observations. We removed an outlying measurements at BJD = 2454541.78467, which had a Signal-to-Noise Ratio (S/N) of $\sim 10$.
Figure \ref{fig:1} shows the RVs, their $\Delta\ln L$ periodogram, best-fit Keplerian model in time and phase, RV residuals, and their $\Delta\ln L$ periodogram. The $\Delta\ln L$ periodogram showed three significant signals at periods of $1893$, $462$, and $301$\,d, as evident from the false alarm probability (FAP, calculated as in \citet{trifon2020b}) being $<0.001$. The peaks at $462$\,d and $301$\,d were found to be the $\sim 1$\,yr aliases of the $1893$\,d period. The most significant peak at $1893$\,d was used as an initial value for the modelling of the RV data.

We generated posterior samples of the parameters using nested sampling. The priors, specified in Table \ref{tab:tab2}, were chosen after trial and error procedure in which we selected the set of prior values for which highest $\Delta\ln\mathcal{Z}$ was obtained. We also generated posterior samples with a linear trend term in the model, but the decreased $\Delta\ln\mathcal{Z}$ of such a model, with respect to the null model (see Table \ref{tab:tab3}), shows that there is no evidence for a linear trend in the data. 

Comparing an eccentric model with a circular one, on the basis of $\ln\mathcal{Z}$ values, shows clear evidence for an eccentric orbit. Upon the subtraction of this signal, no significant peaks remained (Fig. \ref{fig:1}). The $\ln\mathcal{Z}$ values of the different models are given in Table \ref{tab:tab3}. For completeness, we also modeled the original DRS RVs of HD\,103891. Although the periodic signal caused by HD\,103891\,b is evident in these RVs too, the eccentric model has $\Delta\ln\mathcal{Z}=86.50$ relative to the null model, showing that the NZP-corrected SERVAL RVs are better in this case. Moreover, for DRS RVs the best-fit jitter ($\sigma_{\rm pre/post}$) was in the range of $4$--$5$\,m\,s$^{-1}$ and rms of the RV residuals was $4.77$\,m\,s$^{-1}$. Compared with RV jitter of $3$--$4$\,m\,s$^{-1}$ and residuals' rms of $3.96$\,m\,s$^{-1}$ for SERVAL RVs, this indicates a higher precision of the latter.

\begin{figure}[t!]
 \includegraphics[width = 0.95\linewidth]{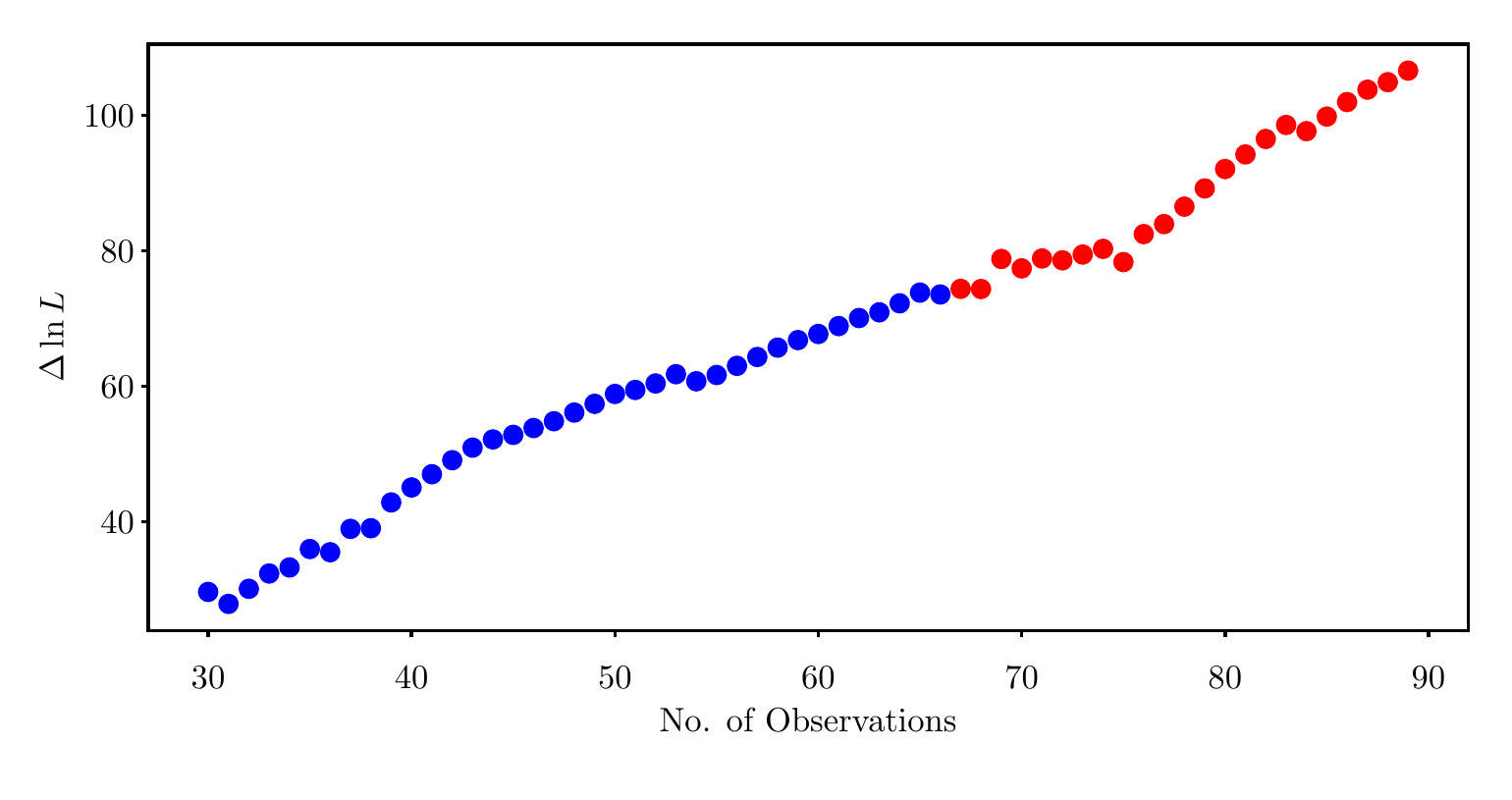}
 \caption{Evolution of the significance of the RV signal for HD~103891 in terms of $\Delta\ln L$. Blue and red dots shows pre and post data. \vspace{-0.3cm}}
 \label{fig:3}
\end{figure}

The posterior PDFs of the the parameters are presented in Fig. \ref{fig:a3}.
We started the coherence analysis with a minimum of 30 data points and re-calculated $\Delta\ln L$ upon the addition of each subsequent RV measurement. The result is plotted in Fig. \ref{fig:3}. The evolution is almost linear in time over the $\sim15$ years of observations, suggesting that the best description for the detected signal is a reflex Keplerian motion of HD\,103891 due to a Jupiter-like planet, HD\,103891\,b.

From the Keplerian parameters, we derived the semi-major axis (a), the minimum mass of the planet ($m \sin{i}$), and the minimum astrometric signature ($\alpha_{\rm min}$) induced by the planet on the host star. These were found to be $3.27$\,au, $1.44$\,M$_{\rm Jup}$, and $58.65$\,$\mu$as, respectively. In addition, we derived the uncertainties of the derived parameters (a, $m \sin{i}$, and $\alpha_{\rm min}$) by generating posterior samples assuming Gaussian distributions for stellar mass and parallax \citep[e.g.,][]{levbarnard2019,Reffert2011}. The best-fit model parameters, derived parameters, and their uncertainties are given in Table \ref{tab:tab4}.
\subsubsection{Spectroscopic activity indicators}
\begin{figure}[t!]
\includegraphics[width = 1.\linewidth]{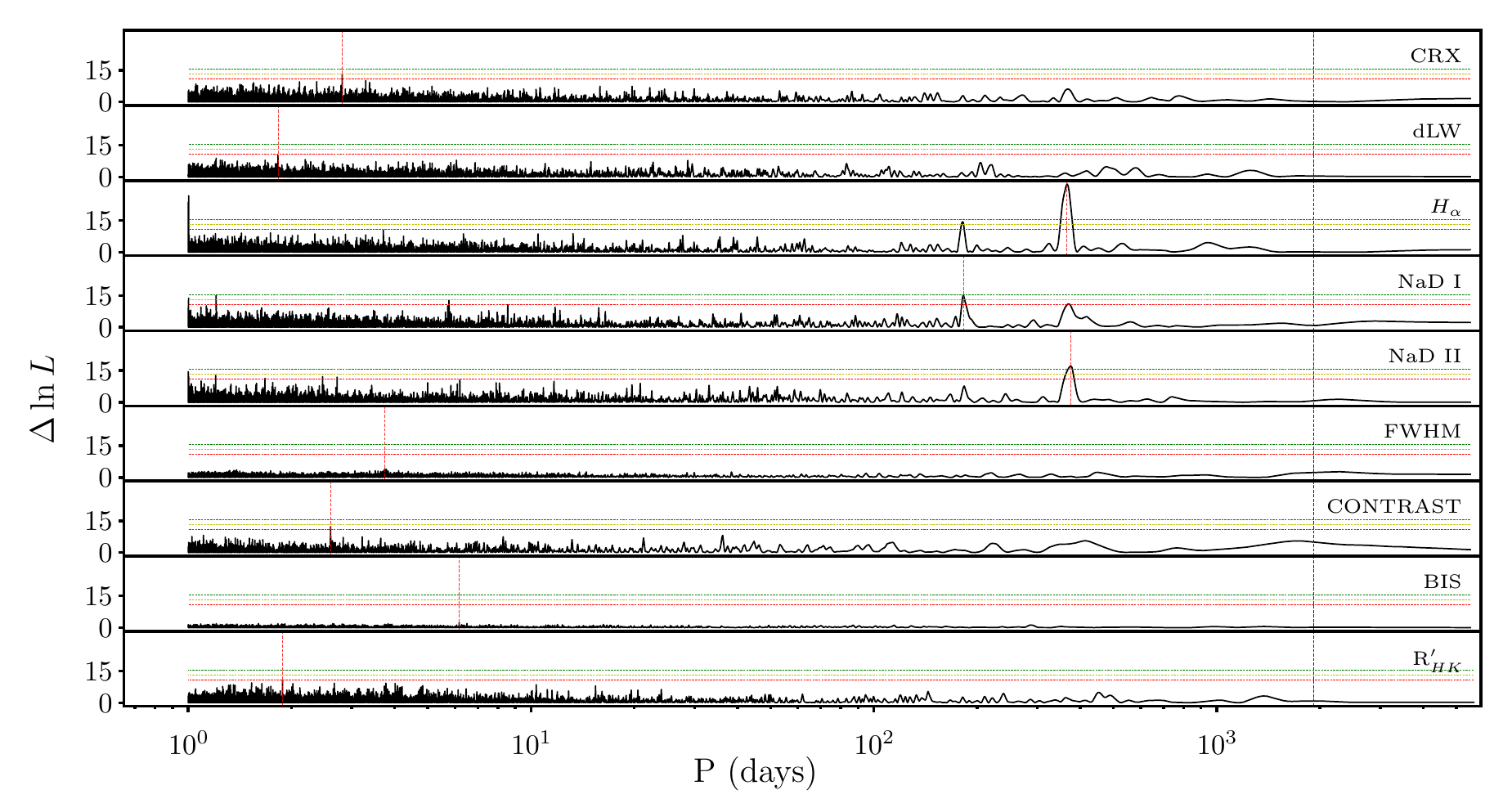}
 \captionsetup{width=.95\linewidth}
\caption{$\Delta\ln L$ periodograms of the spectroscopic activity indicators of HD\,103891. The vertical red lines show the peak period of each signal. The Horizontal lines in each panel mark, from top to bottom, the 0.001, 0.01, and 0.1 FAP values. The blue vertical line shows the period of the proposed planet candidate.}
\label{fig:2}
\end{figure}

Each activity time series as obtained from the {\sc HARPS-RVBank}, was fed to a $\Delta\ln L$ periodogram as shown in Fig. \ref{fig:2}. 
None of the time series show a significant peak at the period of the planet candidate (marked as a blue line in Fig.~\ref{fig:2}). The CRX data does not have any significant long term signals, and the highest peak at $2.8$\,d is only significant at an FAP threshold of 10\%. The periodograms of H$_{\alpha}$ and NaD II show a peak near $365$\,d, and that of NaD I exhibits a peak at $182.4$ days. From window function analysis, these two peaks were found to be aliases. The differential line width (dLW) shows a less significant peak near $1.8$\,d. We found two outliers at BJD$=2454541.78467$ and BJD$=2456051.70645$, possibly due to low S/N observations. 

\onecolumn

\twocolumn
\begin{table}[!ht]
 \centering
 \caption{Parameter priors}
 \setlength{\tabcolsep}{2pt} 
 \renewcommand{\arraystretch}{1.2}
 \begin{tabular}{c c c }
 \hline
 \hline
 Parameter & HD\,103891 & HD\,105779 \\[0.1ex]
 \hline
 \hline
 P\,[d] & $\mathcal{U}$\,(1800, 2000) &$\mathcal{U}$\,(2200, 2700)\\
 K\,[m s$^{-1}$] & $\mathcal{U}$\,(18, 25) & $\mathcal{U}$\,(5, 15) \\
 $e$ & $\mathcal{U}$\,(0,1)&$\mathcal{U}$\,(0,1)\\
 $\omega$\,[deg] & $\mathcal{U}$\,(0, 360)&$\mathcal{U}$\,(0, 360)\\
 t$_{0}$ - 2450000\,[d]& $\mathcal{U}$\,(5024 , 7024) &$\mathcal{U}$\,(5135,7635) \\
 $\gamma_{\rm pre}$\,[m s$^{-1}$] & $\mathcal{U}$\,(-50,50)&$\mathcal{U}$\,(-50,50)\\
 $\gamma_{\rm post}$\,[m s$^{-1}$] & $\mathcal{U}$\,(-50,50)&$\mathcal{U}$\,(-50,50)\\ 
 $\sigma_{\rm pre}$\,[m s$^{-1}$] & $\mathcal{U}$\,(0.0, 20)&$\mathcal{U}$\,(0.0, 20)\\
 $\sigma_{\rm post}$\,[m s$^{-1}$] & $\mathcal{U}$\,(0.0,20)&$\mathcal{U}$\,(0.0,20)\\
 \hline
 \hline
 \end{tabular}
 \label{tab:tab2}
\end{table}


On removing 2 outliers using $3\sigma$ clipping, the location and power of this signal remain unchanged.
\noindent FWHM from the DRS show a low significance peak near $3.7$\,d. We could not find any outliers during visual inspection of FWHM measurements. We modelled this signal with a sinusoidal, removed the fit from the data, and repeated the $\Delta\ln L$ analysis. The residuals contained no significant signal near the proposed $1918$-d period of the planet candidate. We also found 5 outliers in the contrast timeseries, and their removal did not produce any significant peaks. The $\mathrm{R}_{HK}^{'}$ time series periodogram showed a peak around $1.88$\,d, but no long period signals. Overall, the signal at $\sim1918$\,d had no counterpart in spectroscopic activity indices, with or without the removal of outliers.

\subsubsection{Photometric data}
The ASAS project observed the star from HJD $= 2453449.7$ to HJD $=2455057.5$, producing $605$ V-band photometric points. Each data point in the time series was flagged as A, B, C and D based on the quality of observation, with A and B being best. We removed all photometric data flagged as C and D, as well as additional points using $3\sigma$ clipping, and those acquired between HJD $=2452625.8$ and HJD $=2452977.9$ due to possible systematic variations. The remaining $415$ points were fed to a GLS periodogram that showed a significant peak near $400$\,d (Fig. \ref{fig:a1}), which we modelled using a sinusoidal and removed from the data.  The residuals contained no significant peaks. We also used $92$ Hipparcos H-band photometric data, and obtained a non significant peak near $2.2$\,d. TESS observed this star in two sectors (sector 9 and 36). A similar possibly-significant peak was observed in the TESS lightcurve at $\sim2.9$\,d for sector 9 and at $8.9$\,d for sector 36. We could not detect any long term signals in any of the photometric datasets.

\subsection{HD\,105779}
\begin{figure}[h!]
\begin{subfigure}{0.5\textwidth}
  \includegraphics[width=0.9\linewidth]{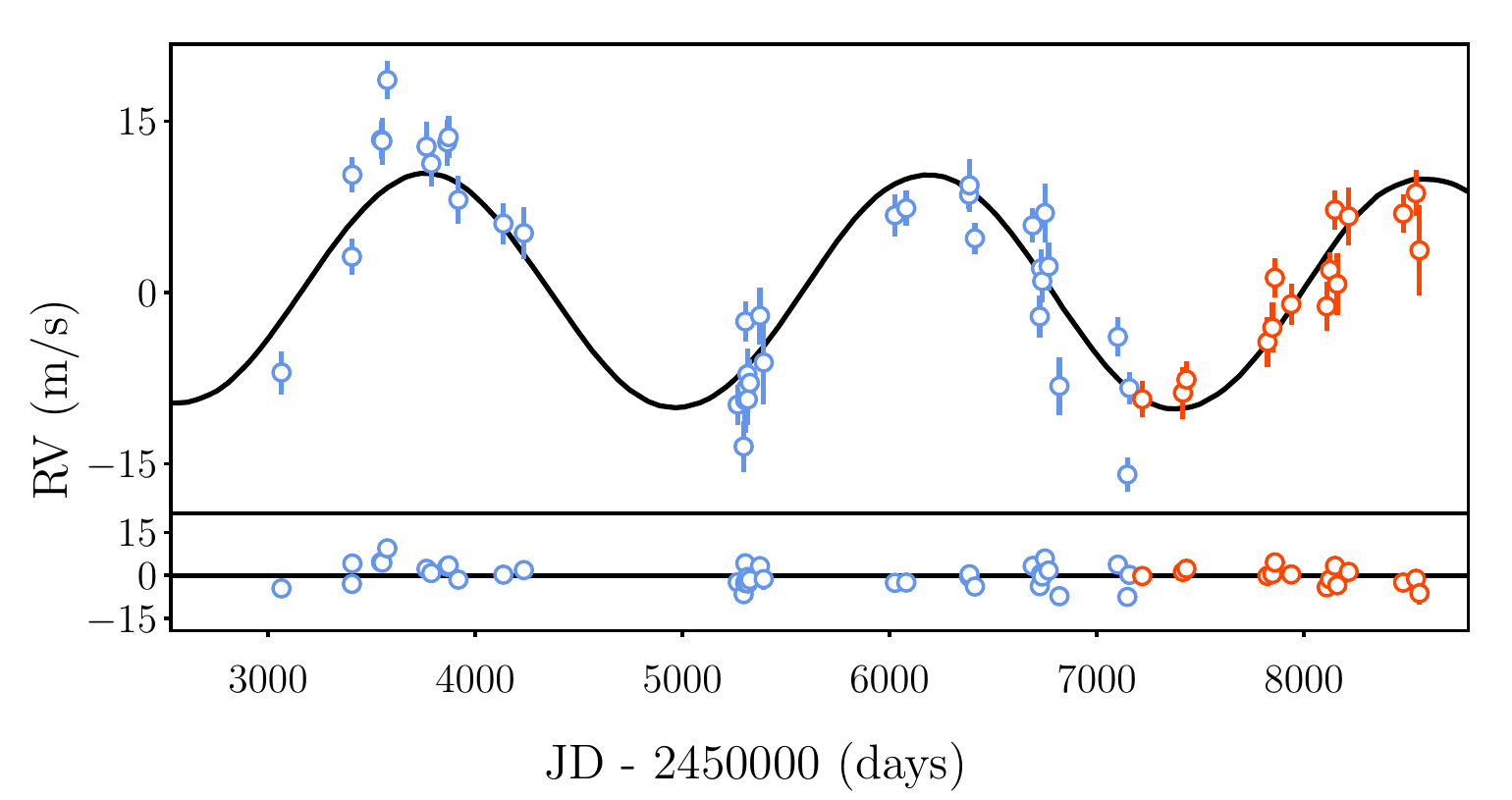}
  \label{fig:subim5}
\end{subfigure}
\hfill
\begin{subfigure}{0.5\textwidth}
  \includegraphics[width=0.9\linewidth]{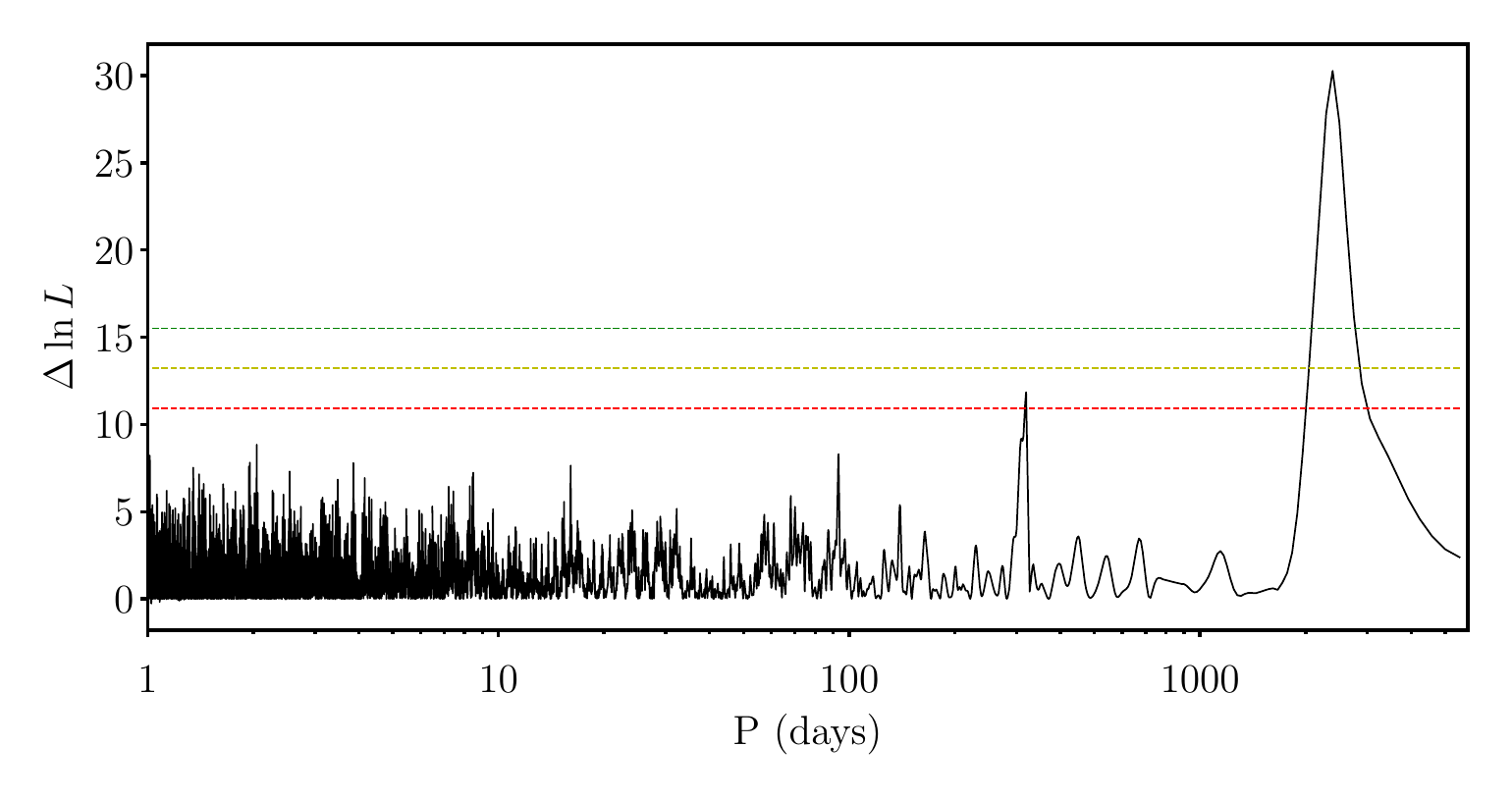}
  \label{fig:subim6}
\end{subfigure}
\hfill
\begin{subfigure}{0.5\textwidth}
  \includegraphics[width=0.9\linewidth]{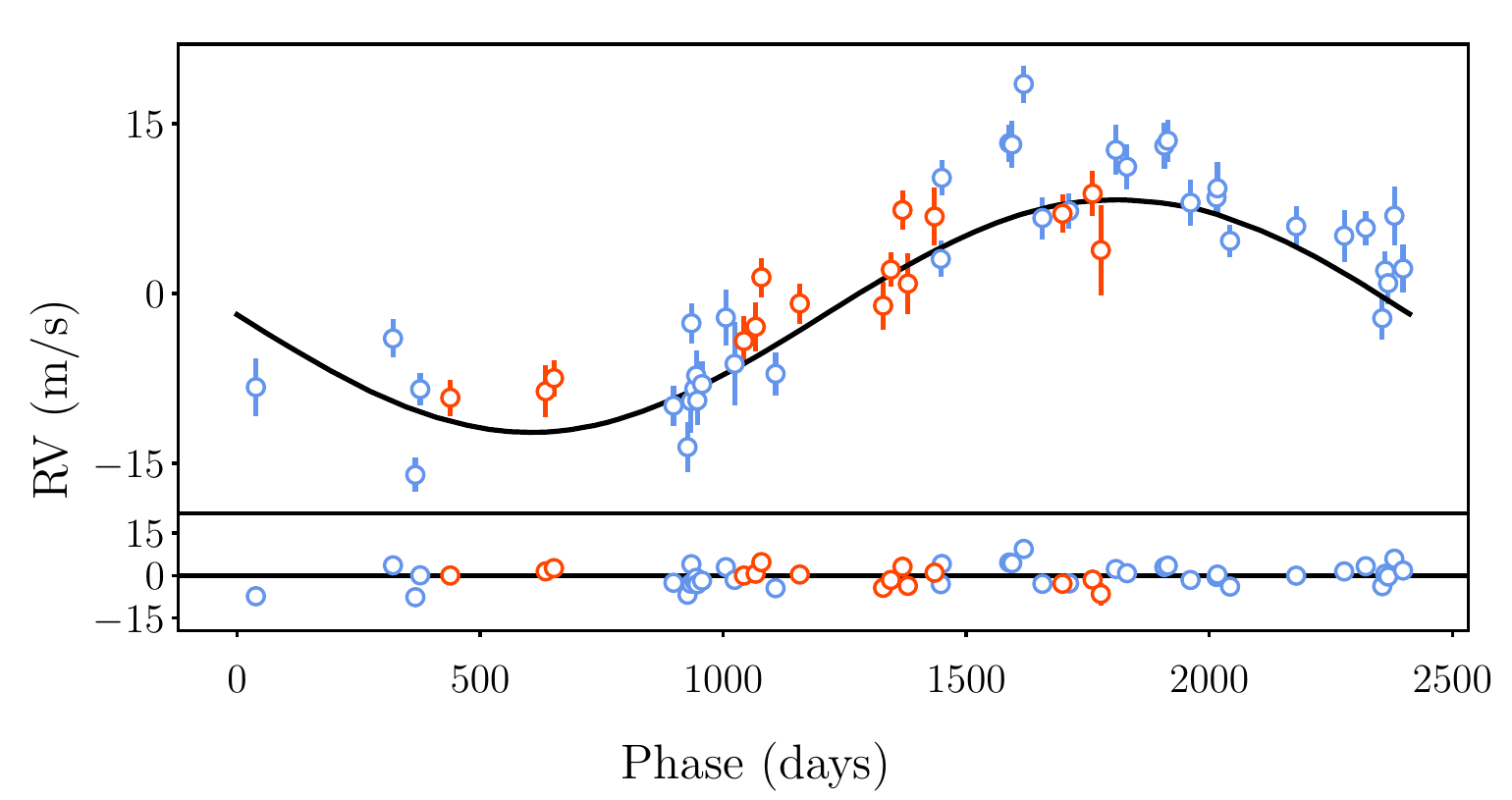}
  \label{fig:subim7}
\end{subfigure}
\hfill
\begin{subfigure}{0.5\textwidth}
  \includegraphics[width=0.9\linewidth]{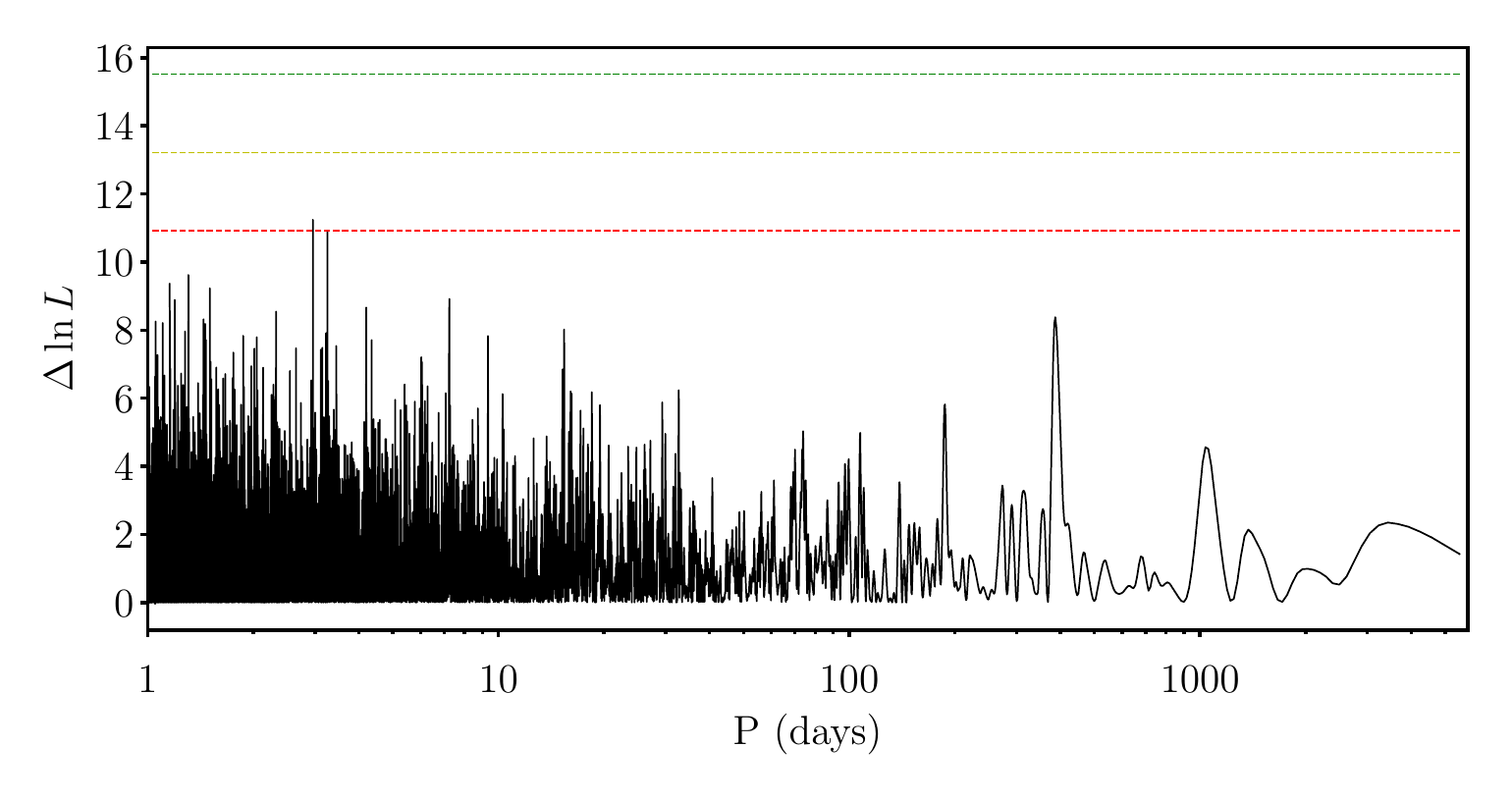}
  \label{fig:subim8}
\end{subfigure}
\captionsetup{width=.95\linewidth}
\caption{Radial velocities and resulting periodograms of HD\,105779. \textit{Top left:} The Keplerian fit to the input RV data. The bottom panel of this figure shows the residual. \textit{Top right:} $\Delta\ln L$ periodogram of input RVs for HD\,105779. The periodogram shows a peak at $2388$\,d. The horizontal dashed lines shows 0.001 (Green), 0.01 (Yellow) and 0.1 (Red) FAP values. \textit{Bottom left:} Data phase folded to pre RVs (Blue) and post RVs (Orange). \textit{Bottom right:} $\Delta\ln L$ periodogram of residuals.}
\label{fig:4}
\end{figure}

HD\,105779 was observed by HARPS from February 2004 to March 2019, producing 53 RV measurements. Figure \ref{fig:4} shows the RVs, their $\Delta\ln L$ periodogram, best-fit Keplerian model in time and phase, RV residuals, and their $\Delta\ln L$ periodogram. The $\Delta\ln L$ periodogram shows a clear, well defined peak at $2388$\,d, with a $\Delta\ln L$ = $31.28$ which is well beyond the FAP threshold lines. The less significant peak at $310$\,d was found to be the $1$\,yr alias of the $2388$\,d signal. Like HD\,103891, the estimation of model parameters was carried out by generating posterior samples using nested sampling. 

We first generated the posterior samples for a Keplerian model, followed by a circular one, by using the priors given in Table \ref{tab:tab2}. In comparison with the null model, the circular orbit resulted in a larger $\Delta\ln\mathcal{Z}$ than the eccentric one (Table \ref{tab:tab3}), which suggests that the orbital eccentricity could not be statistically constrained with the existing data. As with HD\,103891, we also modeled the original DRS RVs of HD\,105779. Relative to the null model, we got $\Delta\ln\mathcal{Z}\lesssim20$ for both the eccentric and circular models. This shows that the detection of HD\,105779\,b with DRS RVs would be marginal. SERVAL RVs, however, present strong evidence for the existence of the planet ($\Delta\ln\mathcal{Z}=35$). The best-fit RV jitter for DRS RVs was in the range of $4$--$6$\,m\,s$^{-1}$ and rms of the RV residuals was $4.79$\,m\,s$^{-1}$. Compared with RV jitter of $2$--$4$\,m\,s$^{-1}$ and residuals' rms of $3.53$\,m\,s$^{-1}$ for SERVAL RVs, this again shows the better precision of the latter.

\begin{table}[!t]
 \centering
 \caption{Model comparison}
 \setlength{\tabcolsep}{2pt} 
 \renewcommand{\arraystretch}{1.2}
 \begin{tabular}{c c c c c }
 \hline
 \hline
 & \multicolumn{2}{c}{HD\,103891} & \multicolumn{2}{c}{HD\,105779}\\
 \hline
 \hline
 Model & ln $ \mathcal{Z}$ & $\Delta$ ln $\mathcal{Z}$&ln $ \mathcal{Z}$ & $\Delta$ ln $\mathcal{Z}$\\[0.1ex]
 \hline
 Null model & -372.08 & 0.0 & -195.06& 0.0\\
 Null model + Linear Trend & -386.26& -14.18 & -206.91&-11.85\\
 Eccentric & -277.76 & 94.32 &-162.11 & 33.49 \\
 Circular & -305.03 & 67.06 & -159.95& 35.11 \\
 \hline
 \hline
 \end{tabular}
 \label{tab:tab3}
\end{table}

\begin{figure}[t!]
 \includegraphics[width = 0.95\linewidth]{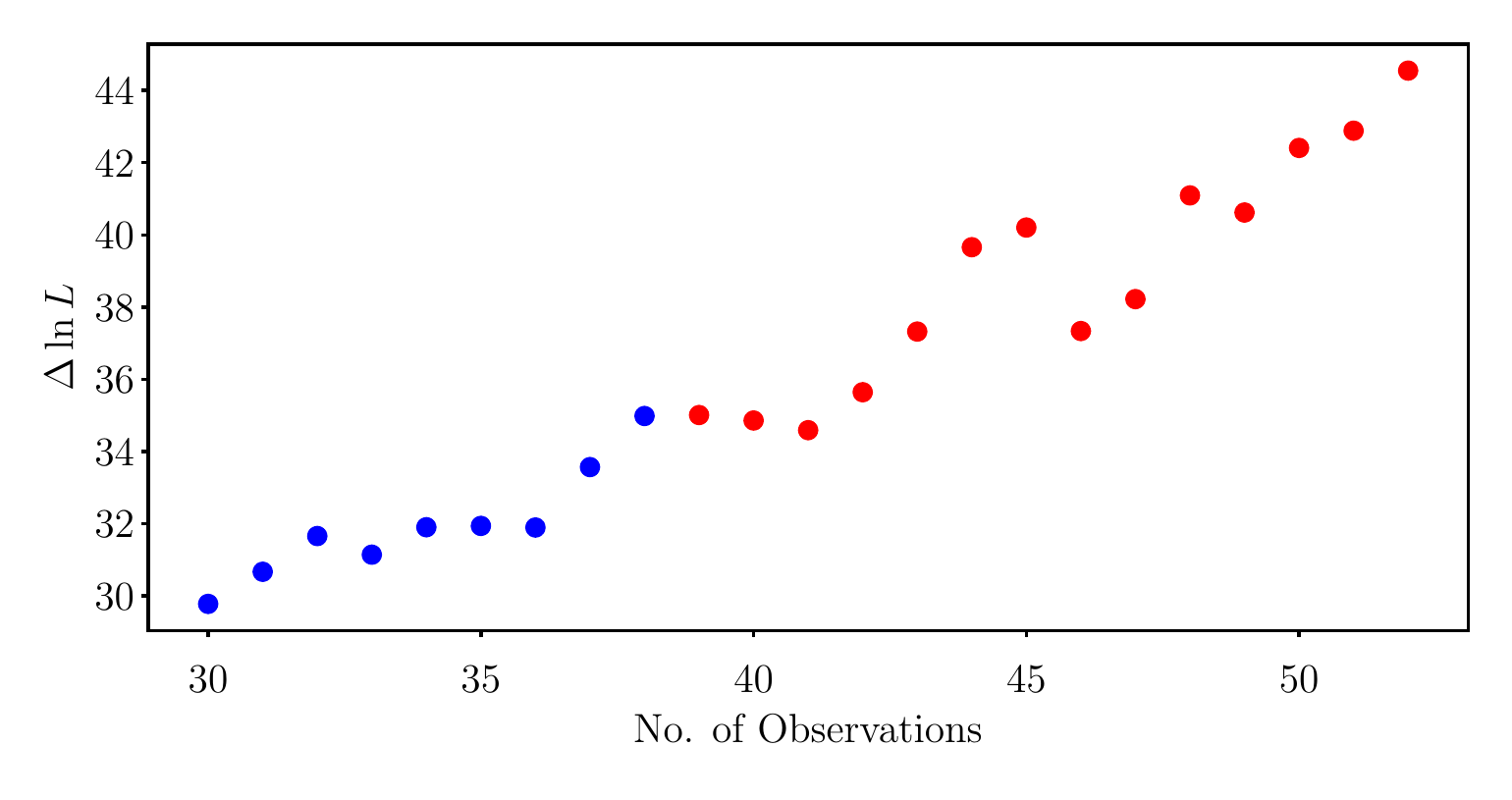}
 \caption{Evolution of the significance of the RV signal for HD\,105779.\vspace{-0.3cm}}
 \label{fig:6}
\end{figure}

The $\Delta\ln L$ evolution (Fig. \ref{fig:6}) also suggests that, with the exception of up to three measurements, the signal is coherent over the observational time span. The posterior distributions of the model parameters is presented in Fig. \ref{fig:a4}. As for HD\,103891, the most probable model parameters, derived parameters, and their uncertainties are given in Table \ref{tab:tab4}. 

\subsubsection{Spectroscopic activity indicators}
\begin{figure}[h!]
\includegraphics[width = 1.\linewidth]{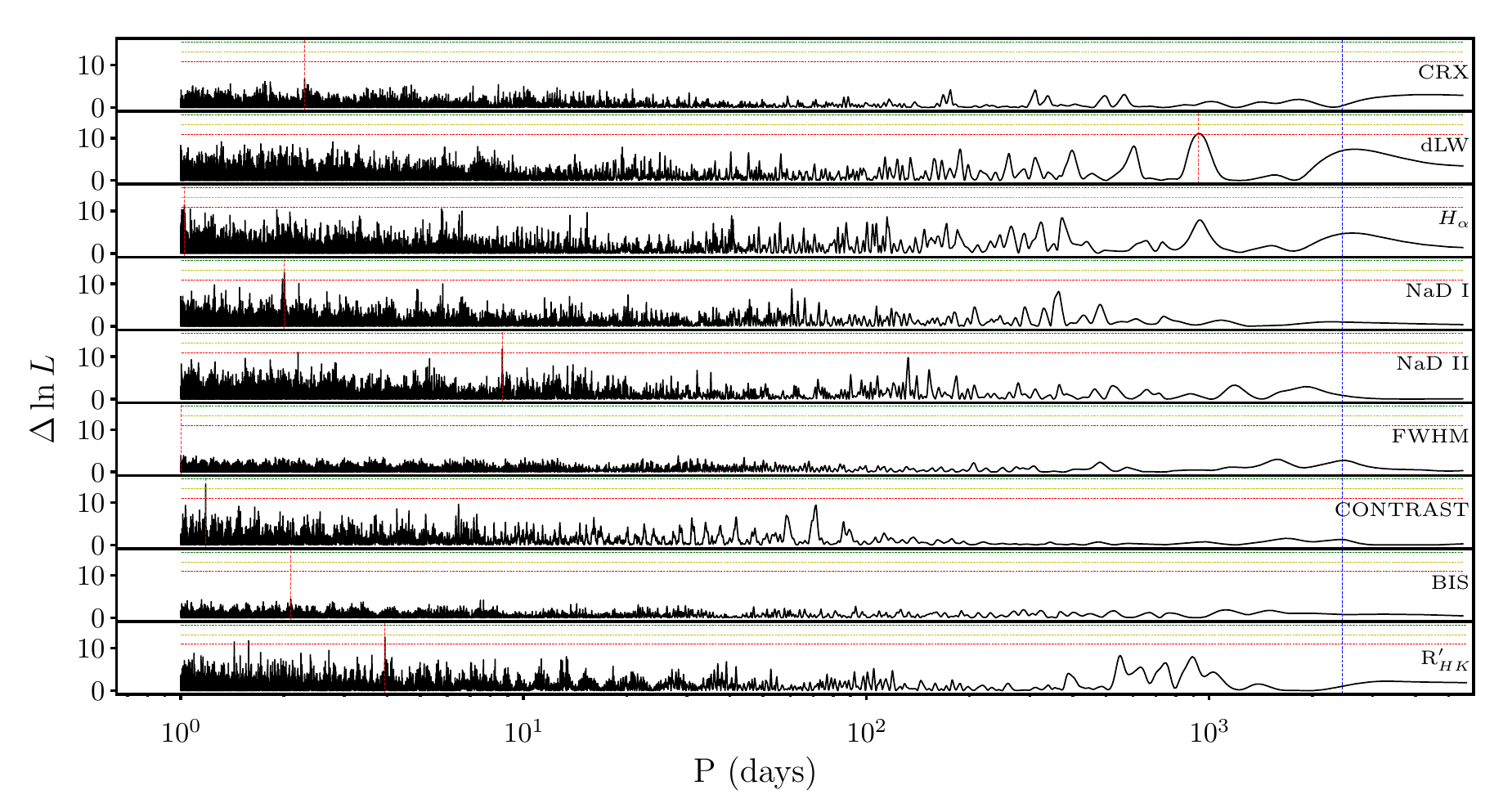}
\captionsetup{width=.95\linewidth}
\caption{$\Delta\ln L$ periodograms of activity indicators for HD\,105779. The vertical red lines show the peak period of the signal in each. The horizontal lines mark the FAP values as in Fig. \ref{fig:5}. The blue vertical line shows the period of the proposed planet.}
\label{fig:5}
\end{figure}

The activity series obtained from HARPS were visualized using $\Delta\ln L$ periodograms in Fig. \ref{fig:5}. The CRX shows an insignificant period at $2.3$\,d, and NaD I exhibits a peak at $2.1$\,d. The BIS and $\mathrm{R}^\prime_{HK}$ show insignificant peaks at $2.1$\,d and $3.9$\,d, respectively. The dLW has a low significance peak at $934$\,d. The FWHM has a power at $1.01$\,d, due to the window function of the dataset. We removed an outlier from FWHM data using $3\sigma$ clipping and found an insignificant period at $45$\,d. Similarly, two outliers were removed from the contrast time series, which yielded an insignificant peak around $2839$\,d. Thus, we could not find any significant signal in activity indices with a period similar to that of the proposed planet.


\onecolumn

\twocolumn

\begin{figure}[t!]
 \includegraphics[width = 0.95\linewidth]{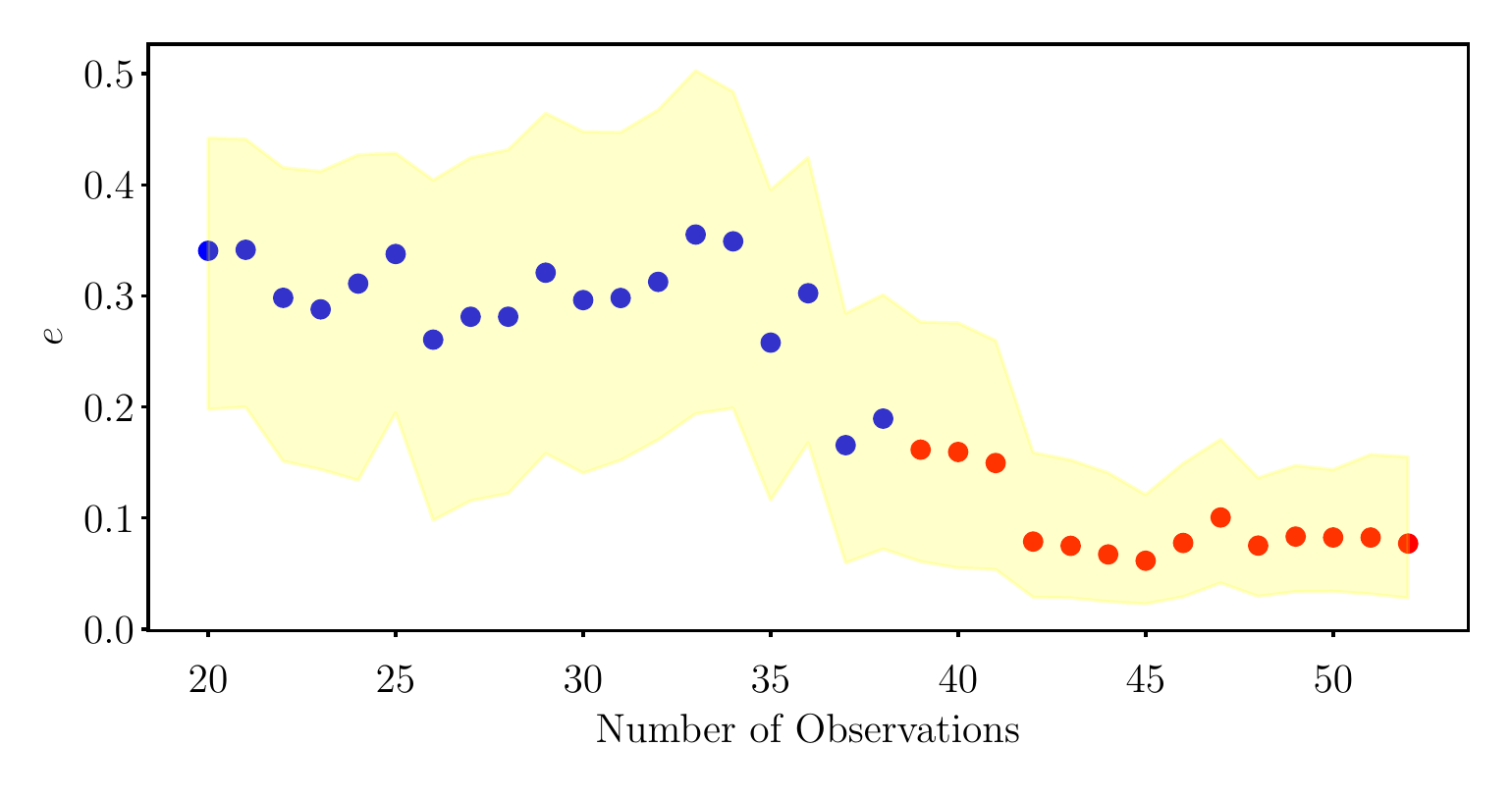}
 \caption{Evolution of the best-fit eccentricity and its uncertainties for the RVs of HD\,105779.\vspace{-0.3cm}}
 \label{fig:7}
\end{figure}

\noindent
\subsubsection{Photometric data}
We checked the ASAS V band photometric data of HD\,105779 obtained from HJD = $2453393.8$ to $2455165.9$, observed in ten visits, and employed $3\sigma$ clipping leaving 470 measurements. Inspection using GLS periodogram did not reveal any significant signal (Fig. \ref{fig:a2}). The GLS periodogram of the Hipparcos photometric magnitudes did not reveal any significant signal either. We also used the available TESS lightcurves (sectors 10 and 36) and analysed them individually, yielding significant peak at short periods.
\section{Discussion}
\label{sec5}
The analysis of public HARPS RV data, as well as of spectroscopic and photometric activity indicators, has resulted in the discovery of two Jupiter-like planets: HD\,103891\,b and HD\,105779\,b. They join the population of cold giant planets
, most of which were detected through long-duration RV monitoring of nearby FGK dwarf stars. It is estimated that $\sim10$\,\% of FGK dwarfs host cold giants \citep[]{Cumming2008}. The population of cold giants is also helpful in understanding the formation of super-Earth planets \citep{cossou2014,zuwu2018,bertram2020}. 

Since the orbital periods of HD\,103891\,b and HD\,105779\,b are larger than any known stellar rotation period of similar spectral-type stars by an order of magnitude \citep[e.g., ][]{2014ApJS..211...24M}, we could exclude rotation as the underlying cause for the RV signals. The absence of significant periodicity of the activity indices at the proposed planetary periods lead us to also exclude magnetic activity cycles as possible causes for the periodic RV variations. We also could not find any significant correlations between RV and activity indices, strengthening this conclusion. The long period and the resulting transit probability ($\sim10^{-3}$) make a photometric confirmation of the candidates unlikely. 
 Interestingly, both stars exhibit a significant astrometric excess noise ($\epsilon_i$) of $<1$\,mas in Gaia EDR3 (Table \ref{tab:tab1}). However, since this is only based on $34$ months of data \citep[][]{lindegren2021}, which is about half of the orbital periods of HD\,103891\,b and HD\,105799\,b, it is too early to use $\epsilon_i$ to derive upper limits on their masses. \citet{kervella2019} used Gaia-Hipparcos proper-motion anomaly (PMa) to find stellar and substellar companions of nearby stars. In an update of that study, \citet[][]{Kervella2021arXiv} could not detect a significant PMa for either HD\,103891 or HD\,105799, neither in Gaia DR2 nor in Gaia EDR3. Nevertheless, they calculated the masses of the unseen companions ($m_{\rm 2}$) that would be required to explain the PMa values, for orbital distances of $3$, $5$, $10$, and $30$\,au. For HD\,103891 they found $m_{\rm 2}(3\,{\rm au}) = 8.0$\,M$_{\rm Jup}$ and $m_{\rm 2}(5\,{\rm au}) = 4.9$\,M$_{\rm Jup}$. For HD\,105799 they found $m_{\rm 2}(3\,{\rm au}) = 2.4$\,M$_{\rm Jup}$ and $m_{\rm 2}(5\,{\rm au}) = 2.0$\,M$_{\rm Jup}$. Unfortunately, uncertainties were only given for $m_{\rm 2}(5\,{\rm au})$ in \citet[][]{Kervella2021arXiv}. Assuming similar uncertainties for $m_{\rm 2}(3\,{\rm au})$ the astrometric $1\sigma$ upper limits on the masses of HD\,103891\,b and HD\,105799\,b are $\sim10$\,M$_{\rm Jup}$ and $\sim4$\,M$_{\rm Jup}$, respectively. These values place them in the realm of massive planets, excluding the possibility of them being brown dwarfs. In the following, we discuss our findings for each of the planets in more detail.

\begin{table}[!t]
 \caption{Median values and $1\sigma$ uncertainties from posterior samples for the companions to HD\,103891 and HD\,105779. 
 }
 \label{tab:tab4}
 \centering
 \setlength{\tabcolsep}{13pt} 
 \renewcommand{\arraystretch}{1.2}
 \begin{threeparttable}
 \begin{tabular}{l c r}
 \hline
 \hline
 Parameters & HD\,103891\,b& HD\,105779\,b\\
 \hline
 \hline
 Null model\\
 \hline
 $\gamma_{\rm pre}$\,[m s$^{-1}$] & 2.11$^{+1.59}_{-1.54}$ &-0.35$^{+1.44}_{-1.41}$ \\
 $\gamma_{\rm post}$\,[m s$^{-1}$] & 2.69$^{+3.39}_{-3.12}$ &0.21$^{+1.58}_{-1.56}$\\
 $\sigma_{\rm pre}$\,[m s$^{-1}$] & 13.04$^{+1.28}_{-1.06}$ &8.85$^{+1.25}_{-0.94}$\\
 $\sigma_{\rm post}$\,[m s$^{-1}$] & 16.15$^{+2.61}_{-2.17}$ &1.75$^{+1.44}_{-1.01}$\\
 \hline
 Planetary\\
 \hline
 P\,[d] & 1919$^{+14}_{-15}$ &2412$^{+55}_{-52}$\\
 K\,[m s$^{-1}$] & 21.12$^{+0.87}_{-0.84}$&10.42$^{+0.97}_{-0.95}$\\
 $e$\tablefootmark{a} & 0.31$^{+0.03}_{-0.03}$& 0 (fixed) \\
 $\omega$\, [deg] & 181.49$^{+8.05}_{-8.17}$& 90 (fixed) \\
 $t_{0}$\tablefootmark{b} - 2450000 [d] & 5245$^{+19}_{-19}$ & 6782$^{+39}_{-38}$\\
 $m \sin{i}$\,[M$_{\rm Jup}$] & 1.44$^{+0.06}_{-0.05}$ & 0.64$^{+0.06}_{-0.06}$\\
 $a$\,[au] & 3.27$^{+0.02}_{-0.01}$ &3.38$^{+0.05}_{-0.05}$\\
 
 $\alpha_{\rm min}$\,[$\mu$as] & 58.65$^{+2.37}_{-2.38}$ &42.15$^{+4.37}_{-4.12}$ \\
 \hline
 Instrumental\\
 \hline
 $\gamma_{\rm pre}$\,[m s$^{-1}$]&-1.89$^{+0.60}_{-0.59}$& -1.95$^{+0.77}_{-0.76}$\\
 $\gamma_{\rm post}$\,[m s$^{-1}$]&4.68$^{+0.93}_{-0.91}$& 0.06$^{+1.26}_{-1.22}$\\
 $\sigma_{\rm pre}$\,[m s$^{-1}$] & 3.95$^{+0.47}_{-0.41}$& 3.51$^{+ 0.65}_{-0.55}$\\
 $\sigma_{\rm post}$\,[m s$^{-1}$] & 3.25$^{+0.74}_{-0.60}$&2.21$^{+1.11}_{-0.90}$\\
 \hline
 Statistical\\
 \hline
median $\sigma_{\rm RV}$\,[m s$^{-1}$] & 1.69&1.86\\
$\sigma(\rm O-C)$\,[m s$^{-1}$] &3.96 &3.53 \\
 $\Delta$ ln $\mathcal{Z}$ & 94.32 & 35.11\\
 \hline
 \hline
 \end{tabular}
\tablefoot{
\tablefoottext{a}{We attribute an upper limit of 0.16 for HD\,105779\,b.}
\tablefoottext{b}{time of inferior conjunction for the circular model of the orbit of HD\,105779.}
}
\end{threeparttable}
\end{table}

\subsection{HD\,103891\,b}
The analysis of HARPS data, composed of 90 RV points, resulted in the detection of a $1916$\,d companion on a well defined eccentric orbit, as indicated by the high value of $\ln \mathcal{Z}$. Even after choosing wide priors on the parameters, the orbital parameters of the planet were well constrained such that the $1\sigma$ uncertainties are almost $1\%$ of the median value. The inclusion of a linear trend within null model was found to be insignificant. The planet's equilibrium temperature, assuming a zero albedo, was found to be $240\pm3$\,K \citep{monta2021}. This value, along with the minimum mass of $1.16$\,M$_{\rm Jup}$, places HD\,103891\,b within the class of cold Jupiters.

The $\log$\,g and $T_\mathrm{eff}$ values place HD\,103891 at the border between main sequence stars and subgiants, and maybe even slightly above it \citep{Gomas}. The mean value of $\mathrm{R}^\prime_{HK}$ shows that the star is less active than the Sun. The activity indices and photometric data (Fig. \ref{fig:2} and \ref{fig:a1}) do not show any indication of stellar rotation. Using the v\,$\sin{i}$ value (Table \ref{tab:tab1}), we derived an upper limit of $40.3$\,d for the rotation period of the star, which fits the expectation for an inactive star of this spectral class. There is no sign of activity at long periods, even when not including a linear trend in our model, leading us to exclude a magnetic activity cycle as a possible cause for the periodic RV variations.
\begin{figure}
 \includegraphics[width = 0.95\linewidth]{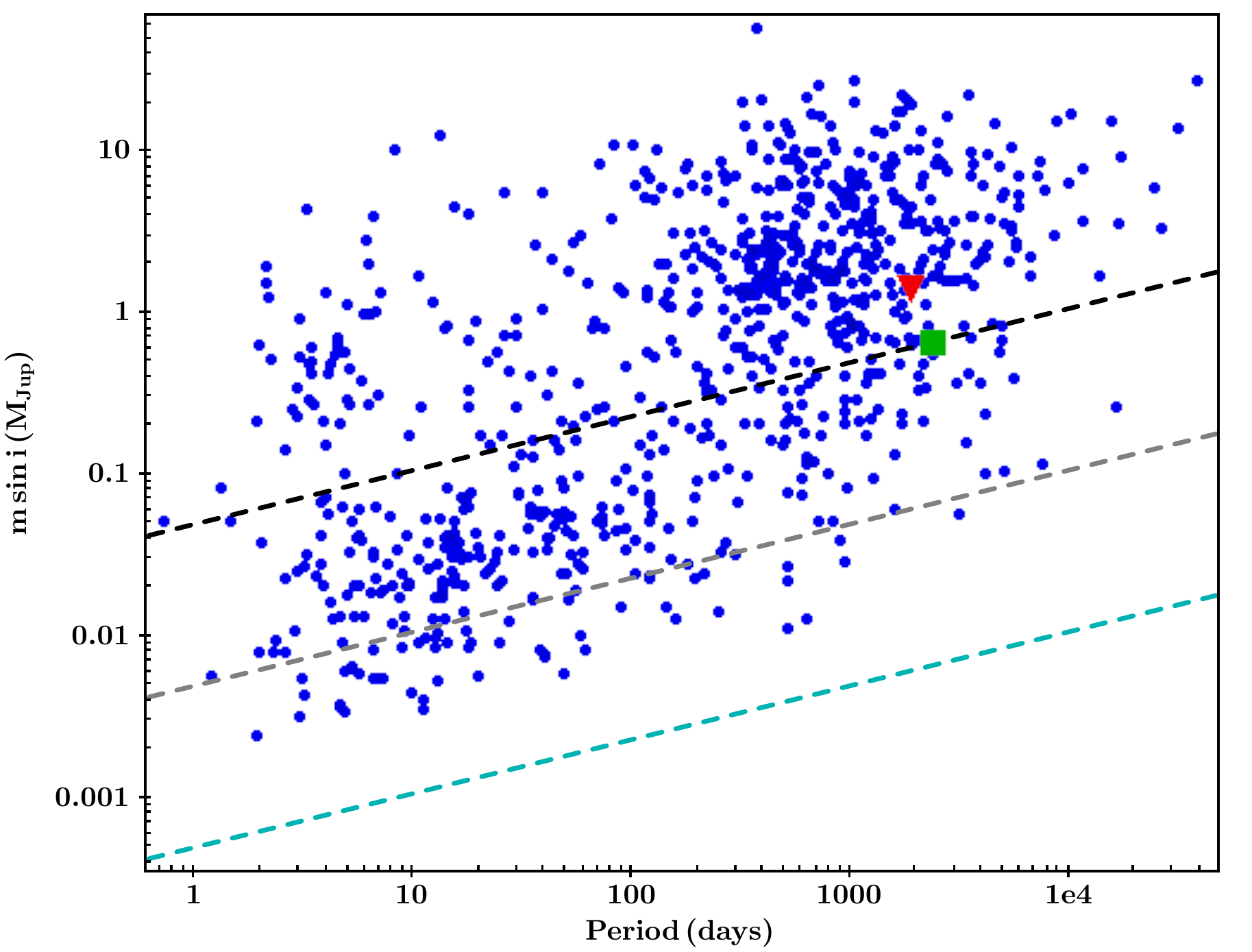}
 \caption{$m\sin{i}$ vs. period distribution of RV detected planets. The red triangle and green square indicate HD\,103891 and HD\,105779, respectively. The black, gray and blue lines indicate detection limits for $1$\,M$_\odot$ host and RV precision of 10, 1, and 0.1 \,m\,s$^{-1}$, respectively.}
 \label{fig:8}
\end{figure}

\subsection{HD\,105779\,b} 
The analysis of $53$ HARPS RV measurements lead to the discovery of a planetary companion with a minimum mass of $0.64\,{\rm M}_{\rm Jup}$ in a $2412$\,d orbit around HD\,105779. Initial analysis of the data using an eccentric model resulted in eccentricity of $0.08^{+0.08}_{-0.05}$. However, the lower number of parameters and the slight increase in $\Delta\ln \mathcal{Z}$ prompted us to favor a circular model, as the best-fit eccentricity may as well be spurious \citep{Lucy1971}. We adopt the value of $0.16$ as the $1\sigma$ upper limit of the orbit eccentricity.

To check whether additional HARPS observations can place better constrains on eccentricity, we analyzed the evolution of the derived eccentricity for HD 105779 b, like we analyzed the evolution of likelihood (Fig. \ref{fig:7}). We found a decay of the derived eccentricity from $0.3\pm0.2$, when considering only the first $\sim30$ RVs, to $0.08^{+0.08}_{-0.05}$, when considering the full RV dataset. This evolution is a typical result of eccentricity bias at low S/N (K$/\sigma(\rm O-C)$) orbital fits \citep{Lucy1971}. The bias is also the reason why the lower eccentricity bound cannot be trusted. Moreover, the estimated eccentricity flattens out with time, so that the last $11$ RV measurements, performed in the years $2016$--$2019$, did not change it. Therefore, few more years of HARPS measurements with the same precision and cadence will not enable a further constraint on eccentricity.

The host star is very similar to the Sun, except it is slightly smaller, older, and of lower metallicity. Orbiting it at a distance of $3.38$\,au, the Jupiter-like planet has an equilibrium temperature of $147\pm2$\,K. 
Although the upper limit for the rotation period of the star ($23.8$\,d) is slightly smaller than that of the Sun, the activity indices and photometric data were unable to provide any hint of the actual rotational period or any long term modulations of the stellar flux (Figures \ref{fig:4} and \ref{fig:a4}).

\subsection{The two planets in context}
\begin{figure}
  \includegraphics[width = 0.95\linewidth]{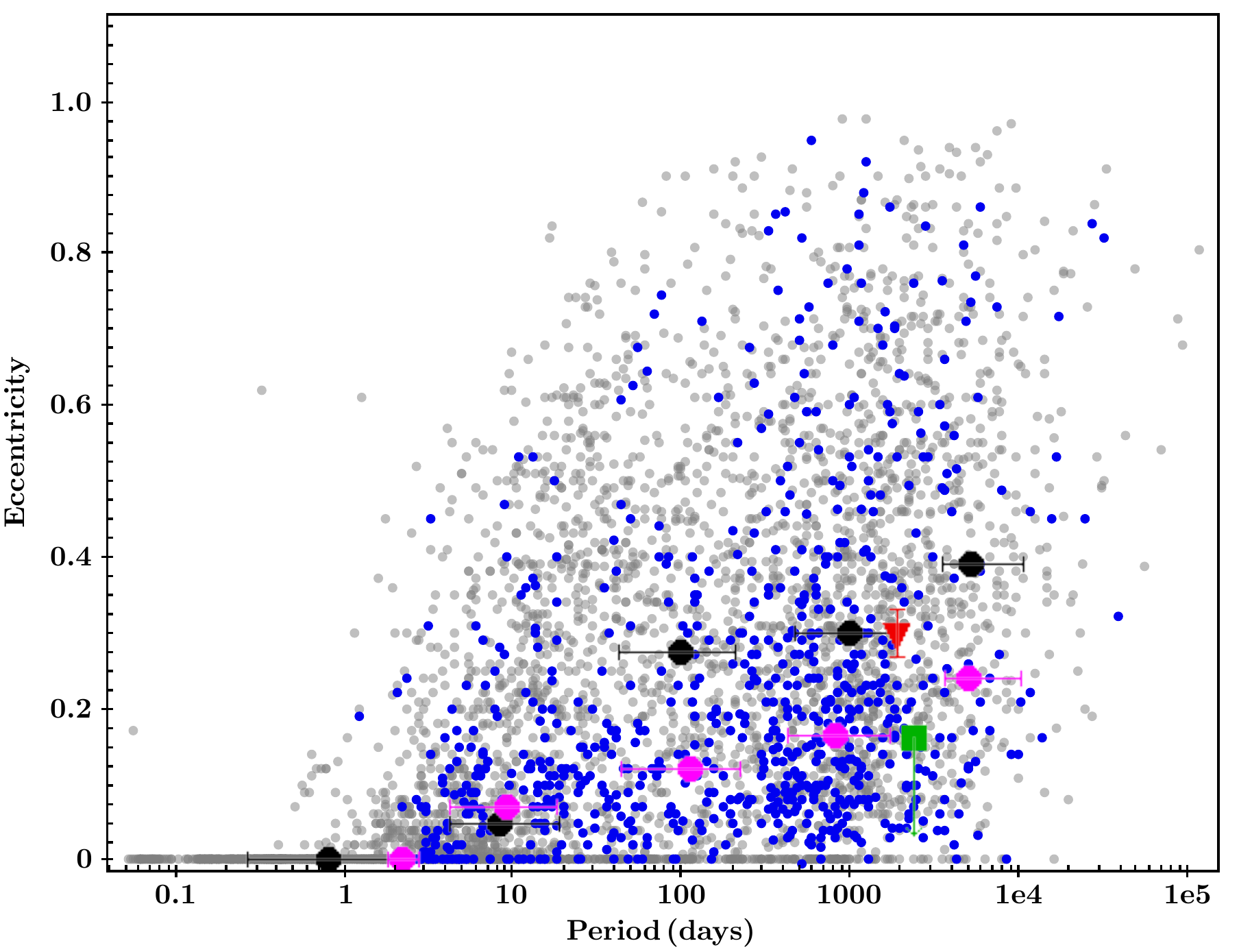}
  \caption{Eccentricity vs. period of binary stars and RV planets. Gray dots correspond to binary stars \citep{sb9} and blue dots to RV planets. The red triangle and green square indicate HD\,103891 and HD\,105779, respectively. Black and pink circles represent the median values of eccentricities in the range $<3$\,d, $3$--$30$\,d, $30$--$300$\,d, $300$--$3\,000$\,d and $>3\,000$\,d for binary stars and planets, respectively.}
  \label{fig:9}
\end{figure}

Figure \ref{fig:8} shows the period vs minimum mass of confirmed planets, discovered using the RV method. HD\,103891\,b and HD\,105779\,b fall well within the distribution and current detection limits. From the eccentricity-period plots shown in Fig. \ref{fig:9}, it can be seen that the eccentricity of the two planets fall on opposite sides of the median value for planets with periods of $>1\,000$\,d. Comparing the eccentricity distribution of planets to that of binary stars \citep{sb9} shows that long-period binaries have on-average slightly higher eccentricities than massive planets. \citet{winfab2015} showed that single planet systems around metal poor stars with long period circular orbits are rare. Although this suggests that HD\,105779 may have additional planetary companions, we could not find them in the existing data. The minimum astrometric signatures induced by the two Jupiter-like planets ($40$--$60\,\mu$as) place them in an interesting parameter space. While currently it is probably impossible to derive their actual masses with astrometry \citep[e.g.,][]{Kervella2021arXiv,Feng2021,Li2021arXiv}, it may become possible with the final Gaia data release \citep[][]{lindegren2021}.

\section{Summary and conclusions}
\label{sec6}
We report the detection of two cold Jupiter-like planets orbiting the slightly evolved F8V star HD\,103891 and the Solar-like star HD\,105779, with periods longer than five years. The orbital parameters of the proposed planets were well constrained from the posterior distribution, generated using nested sampling.

The activity indicators 
were analysed thoroughly, resulting in no significant peaks at the planetary periods. Photometric data from ASAS, Hipparcos and TESS do not exhibit signatures of stellar rotation or activity cycles. The $\mathrm{R}_{HK}^{'}$ values, upper limits on v\,$\sin{i}$, and the lack of rotational modulation indicate low magnetic activity of the host stars, which is in agreement with their estimated age.
Solar-like stars of similar age and activity state typically exhibit activity-related RV variability on the level of $\sim3$\,m\,s$^{-1}$ \citep[e.g.][]{2018A&A...616A.155L}, much lower than the RV variations exhibited by the targets examined here. We conclude that the radial velocity signals of the two stars can only be explained by Jupiter-like companions.

Characterizing the population of cold giant planets around nearby stars ($d<100$\,pc) is timely and important. Not only their true masses could be constrained with astrometric measurements, some of them may even become accessible to direct characterization in reflected light with upcoming instruments, such as the Nancy Grace Roman Space Telescope \citep{Carrion-Gonzalez2021}. Combining results from \textit{Kepler}, RV surveys, and direct imaging campaigns of nearby stars, \citet{Fernandes2019} estimated a giant-planet occurrence rate of $\sim27$\,\% and suggested that there might be a turnover in their occurrence near the snow line of Sun-like stars. Recently, \citet{Fulton2021} extended these findings to beyond the ice line of FGKM stars by detailed statistical analysis of the California legacy RV survey \citep[see also ][]{rosenthal2021}.

The ubiquity of cold giants is also important for understanding the population of inner super-Earth planets. For instance, \citep{zuwu2018} found a probability of $\sim 90$\,\% for having a super-Earth in systems having a cold Jupiter. On the other hand, \citet{bertram2020} used dynamical model simulations to derive an occurrence rate of $30$--$40$\,\% for inner super-Earths. Their simulations found that super-Earths form much before the formation of gas giants. These results contradict the findings of \citet{izidoro2015}, who proposed that gas giants might have been acting as barriers to inner migration of super-Earths. According to \citet{cossou2014}, super-Earths are distant planets that failed to become gas giants. Since our solar system lacks super-Earth planets, studies on the correlations between super Earths and cold giants would help understanding the uniqueness of our Solar system, and may also help in directing observational strategies of future RV surveys.


\begin{acknowledgements}
Based on observations made with ESO Telescopes at the La Silla Paranal Observatory under programme IDs 072.C-0488, 085.C-0019, 087.C-0831, 089.C-0732, 090.C-0421, 091.C-0034, 092.C-0721, 093.C-0409, 095.C-0551, 096.C-0460, 098.C-0366, 099.C-0458, 0100.C-0097, 0101.C-0379, 0102.C-0558, 0103.C-0432, and 183.C-0972. This research is based in part on data obtained by ESA’s Hipparcos satellite, and has made use of the ADS (NASA) and SIMBAD (CDS) services. This paper includes data collected by the TESS mission, which are publicly available from the Mikulski Archive for Space Telescopes (MAST). Trifon Trifonov acknowledge the support by the DFG Research Unit FOR 2544 "Blue Planets around Red Stars" project No. KU 3625/2-1. Funding for the TESS mission is provided by NASA’s Science Mission directorate. We acknowledge the use of \texttt{corner} \citep[][]{corner}, a python tool for the visualization of posterior samples. 
\end{acknowledgements}

\bibliographystyle{aa}
\bibliography{sreenivas.bib}

\begin{appendix}
\onecolumn
\section{Additional plots}

\subsection{Photometry}
\begin{figure}[!ht]
\begin{subfigure}{0.5\textwidth}
  \includegraphics[width=0.9\linewidth]{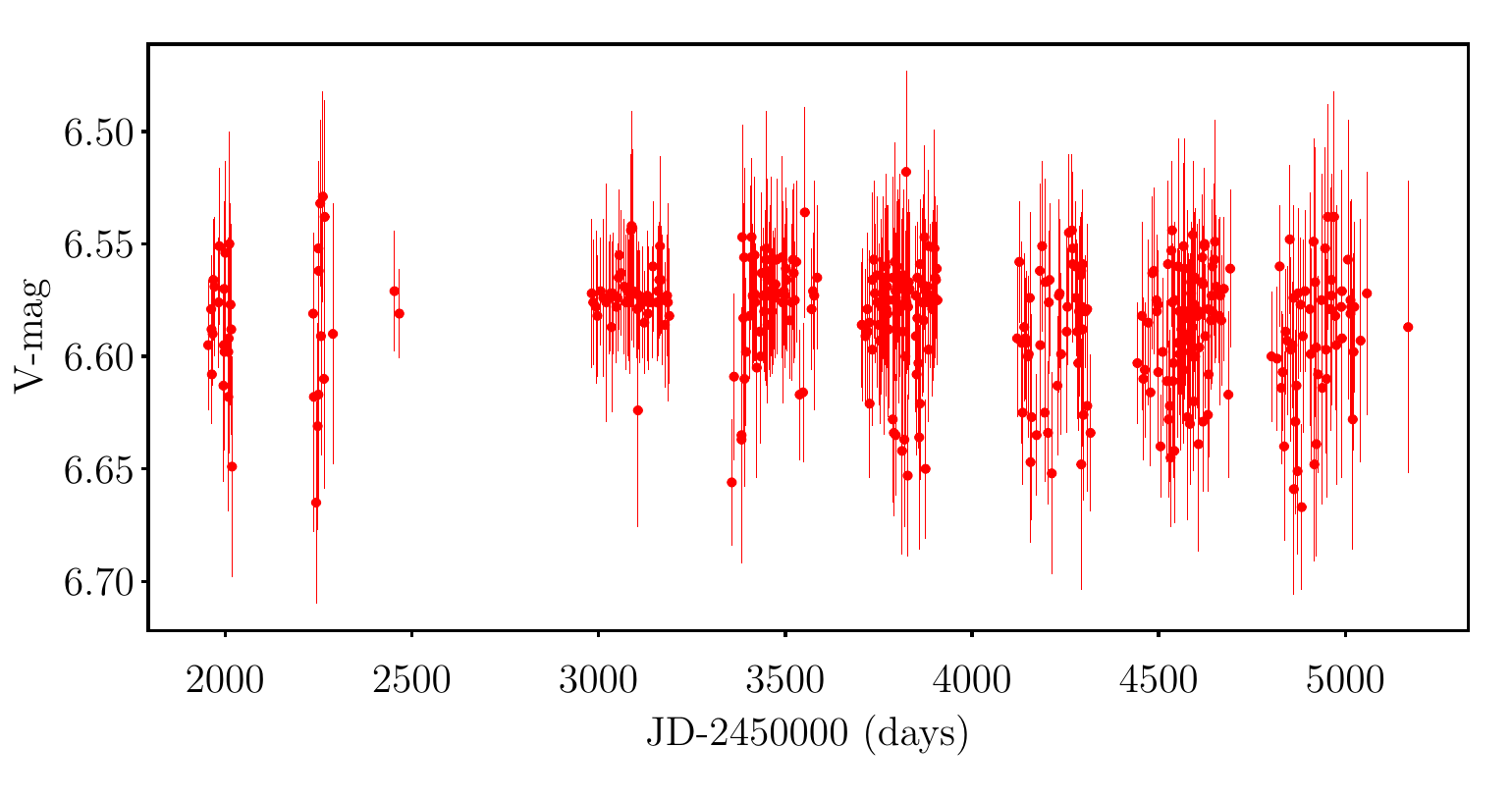}
\end{subfigure}
\begin{subfigure}{0.5\textwidth}
  \includegraphics[width=0.9\linewidth]{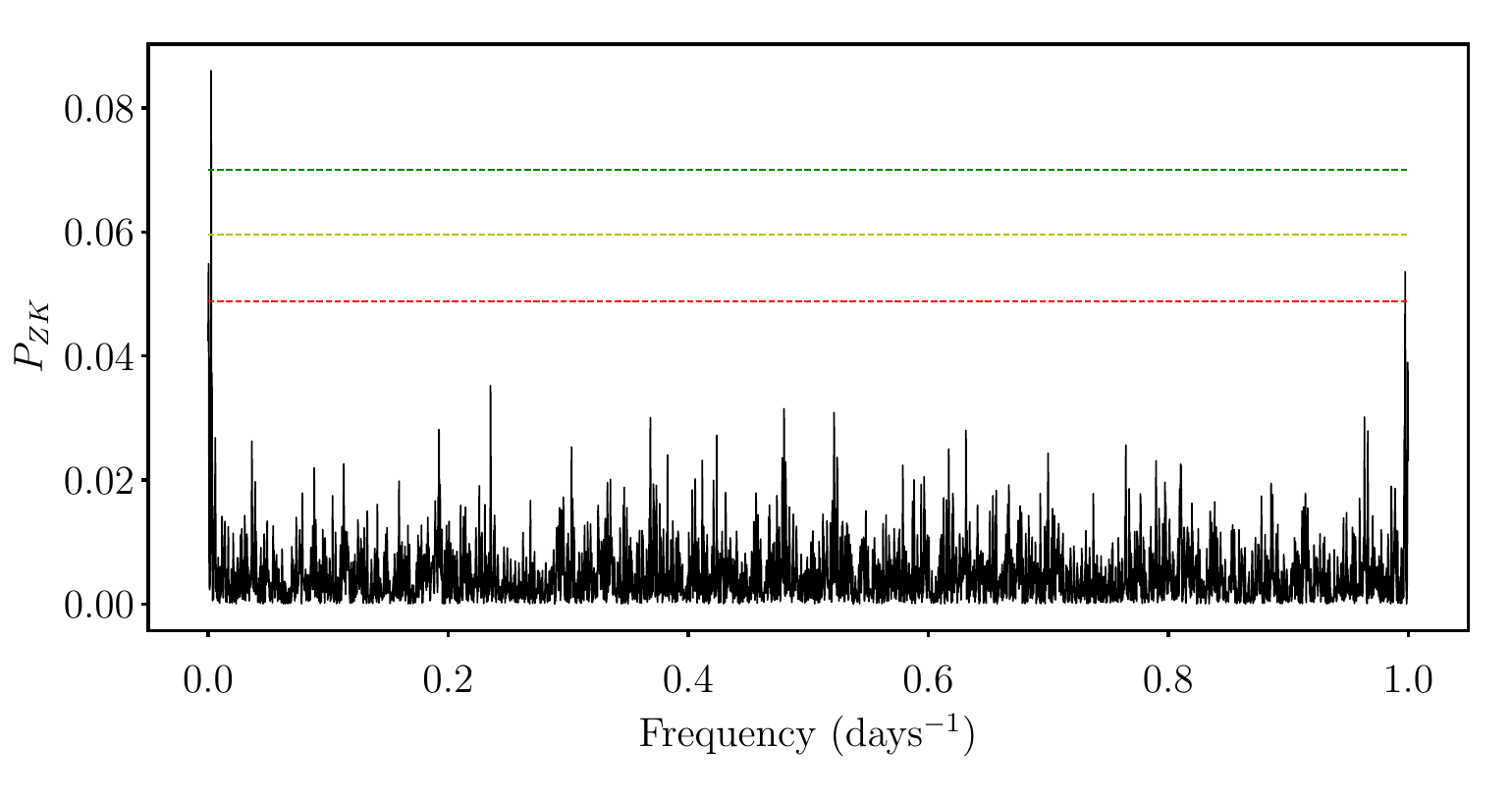}
\end{subfigure}

\begin{subfigure}{0.5\textwidth}
  \includegraphics[width=0.9\linewidth]{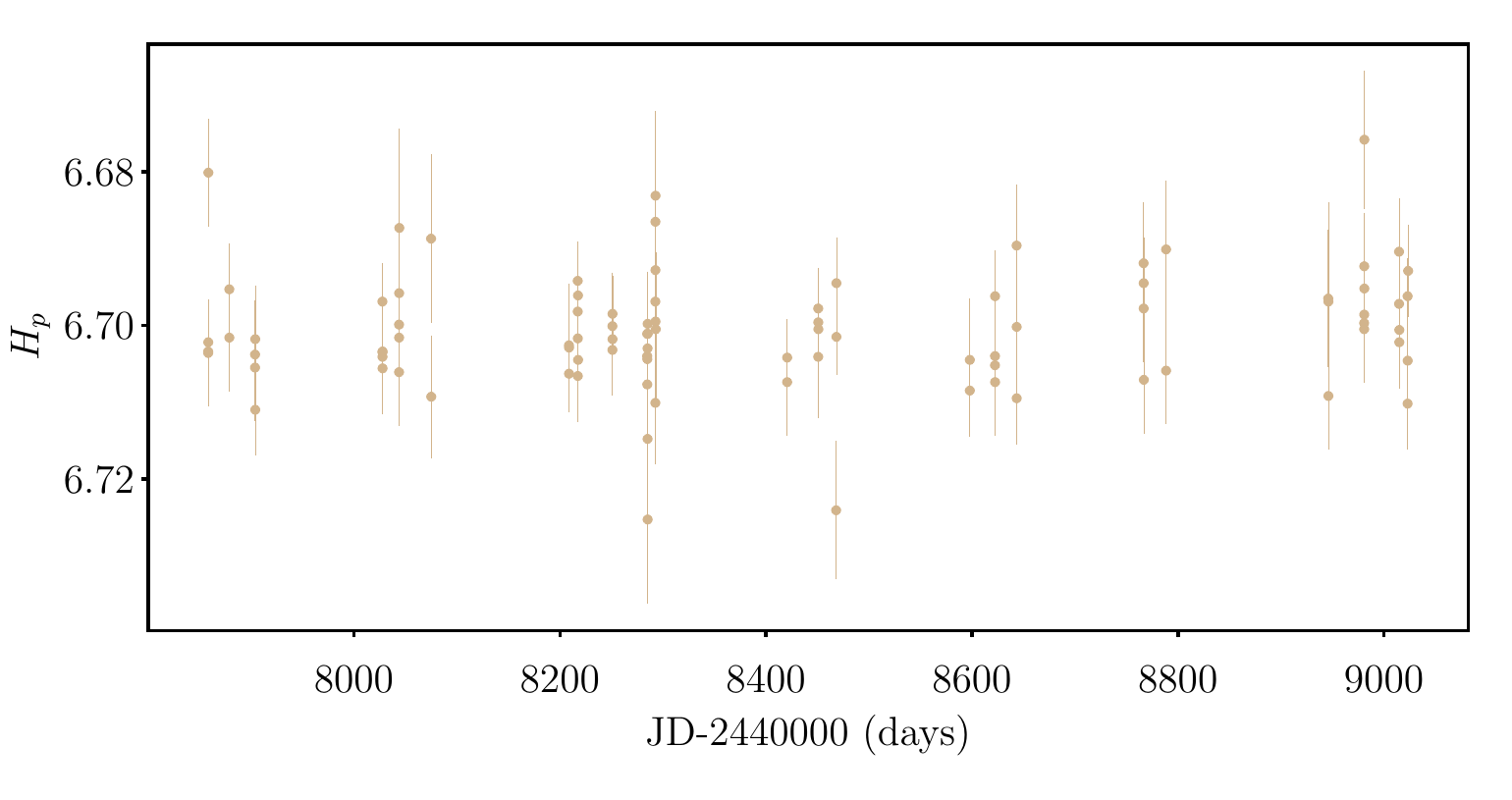}
\end{subfigure}
\begin{subfigure}{0.5\textwidth}
  \includegraphics[width=0.9\linewidth]{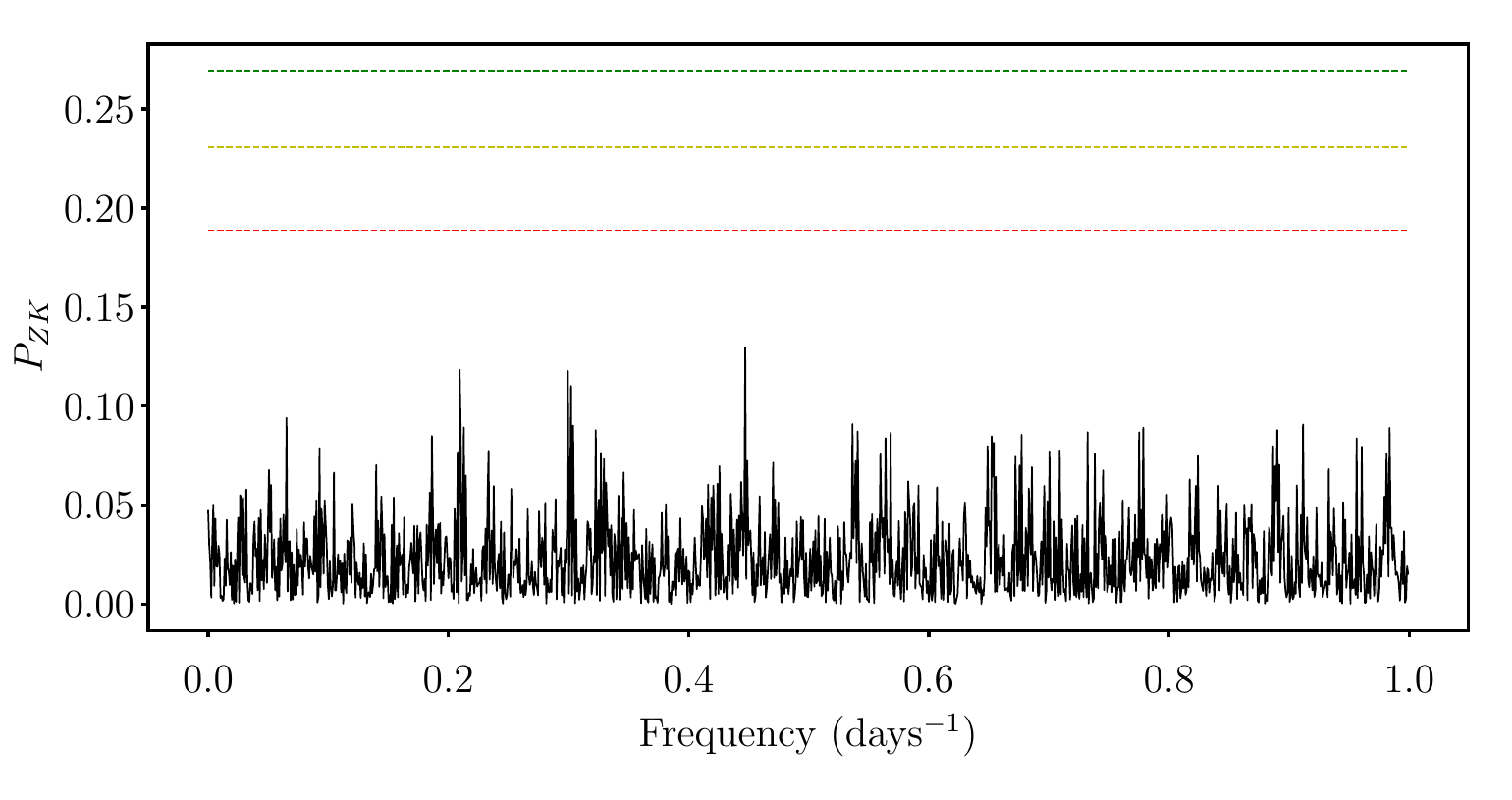}
\end{subfigure}

\begin{subfigure}{0.5\textwidth}
  \includegraphics[width=0.9\linewidth]{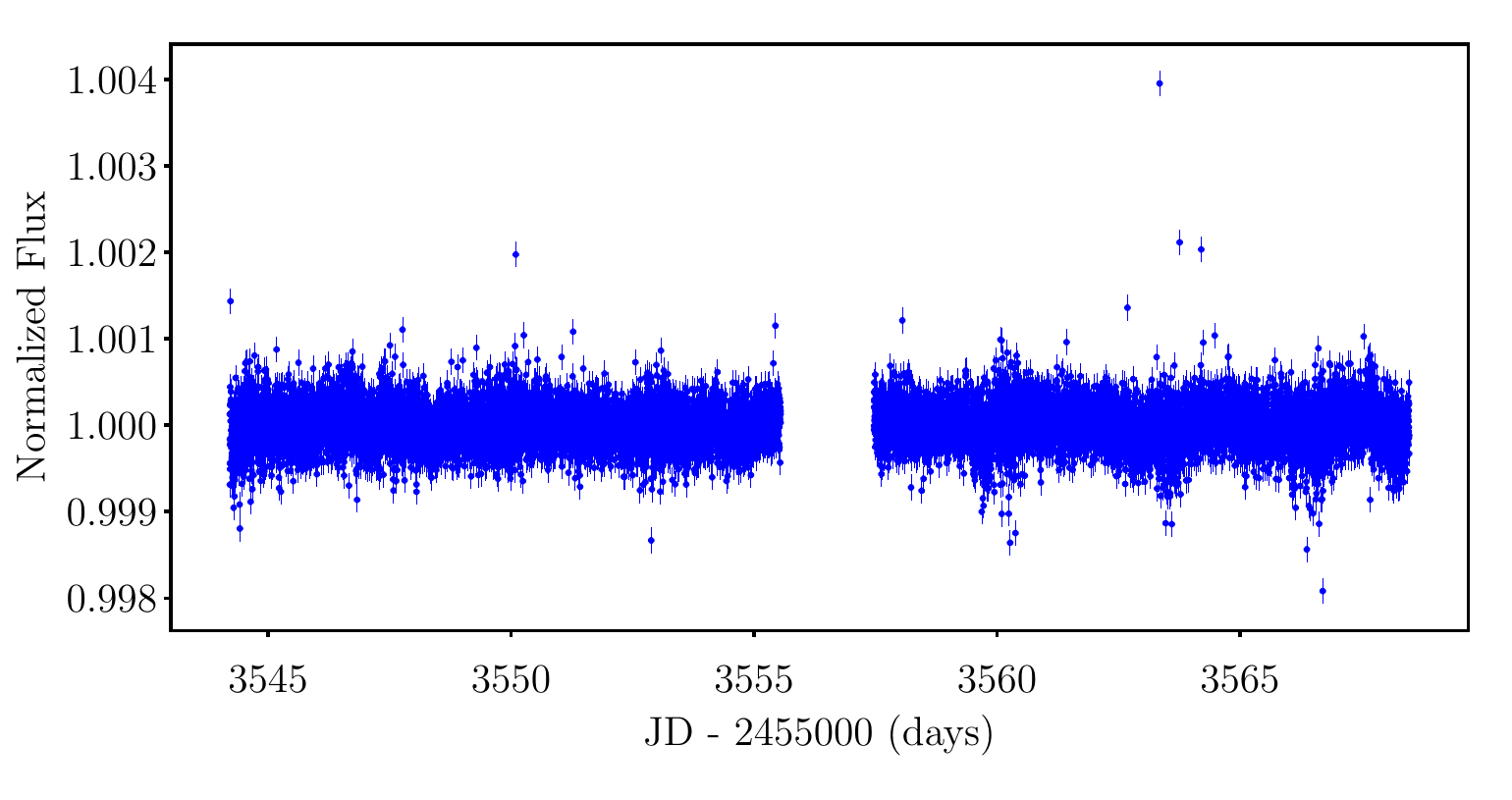}
\end{subfigure}
\begin{subfigure}{0.5\textwidth}
  \includegraphics[width=0.9\linewidth]{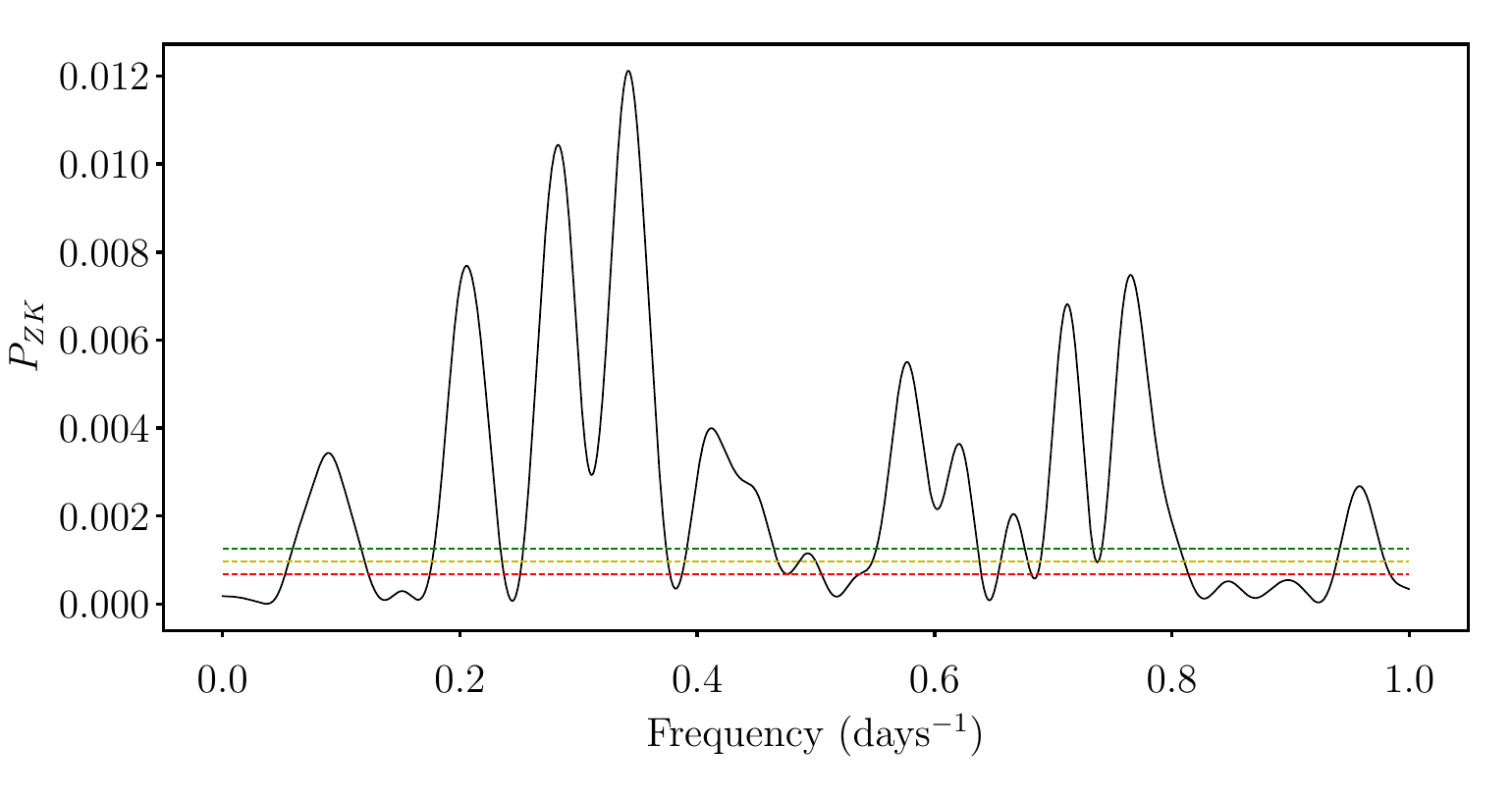}
\end{subfigure}

\begin{subfigure}{0.5\textwidth}
  \includegraphics[width=0.9\linewidth]{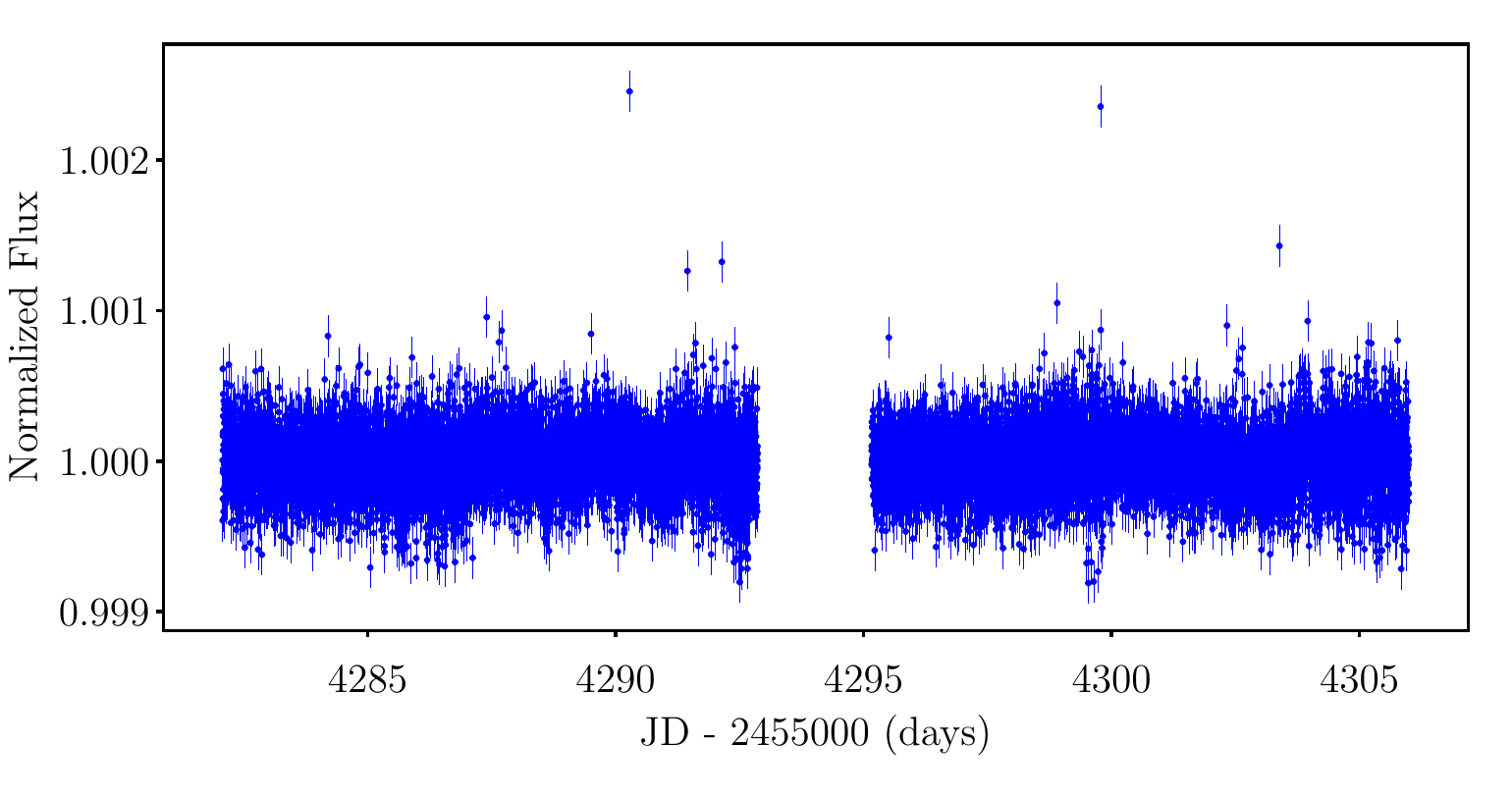}
\end{subfigure}
\begin{subfigure}{0.5\textwidth}
  \includegraphics[width=0.9\linewidth]{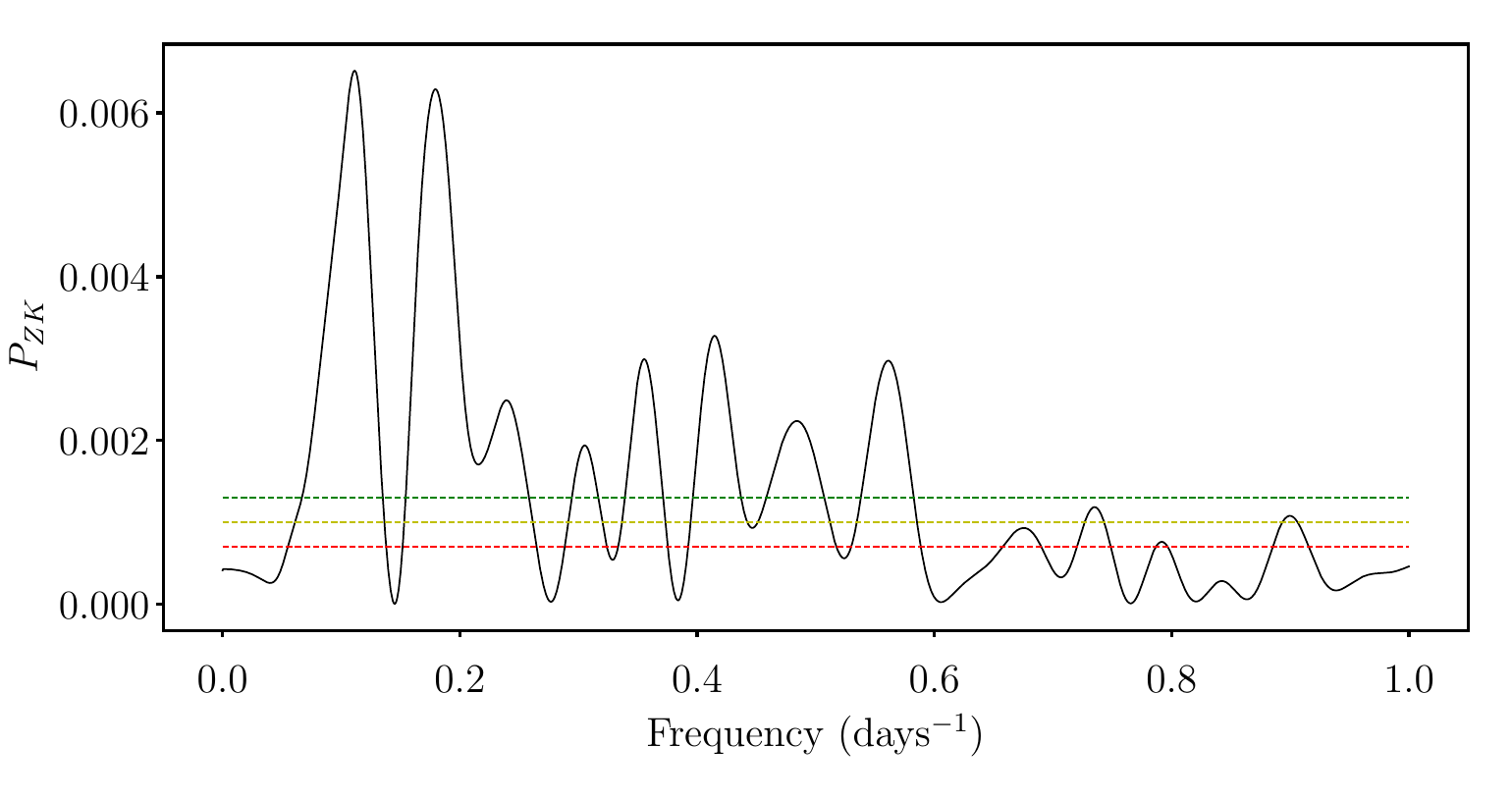}
\end{subfigure}
\caption{Photometry analysis for HD\,103891. \textit{Row 1, Left:} Input time series of ASAS data. \textit{Row 1, Right:} GLS periodogram of ASAS data. The periodogram has a peak at $\sim400$\,d. \textit{Row 2, Right:} Input Hipparcos data. \textit{Row 2, Right:} GLS periodogram of Hipparcos data shows a non significant peak at $\sim2.2$\,d. \textit{Row 3, Left:} Input TESS data for sector 9. \textit{Row 3, Right:} GLS periodogram of TESS flux values shows a peak at $\sim2.9$\,d. \textit{Row 4, Left:} TESS data for sector 36. \textit{Row 4, Right:} GLS periodogram of TESS flux values shows a peak at $\sim8.9$\,d. 
}
\label{fig:a1}
\end{figure}

\begin{figure}[h!]
\begin{subfigure}{0.5\textwidth}
  \includegraphics[width=0.9\linewidth]{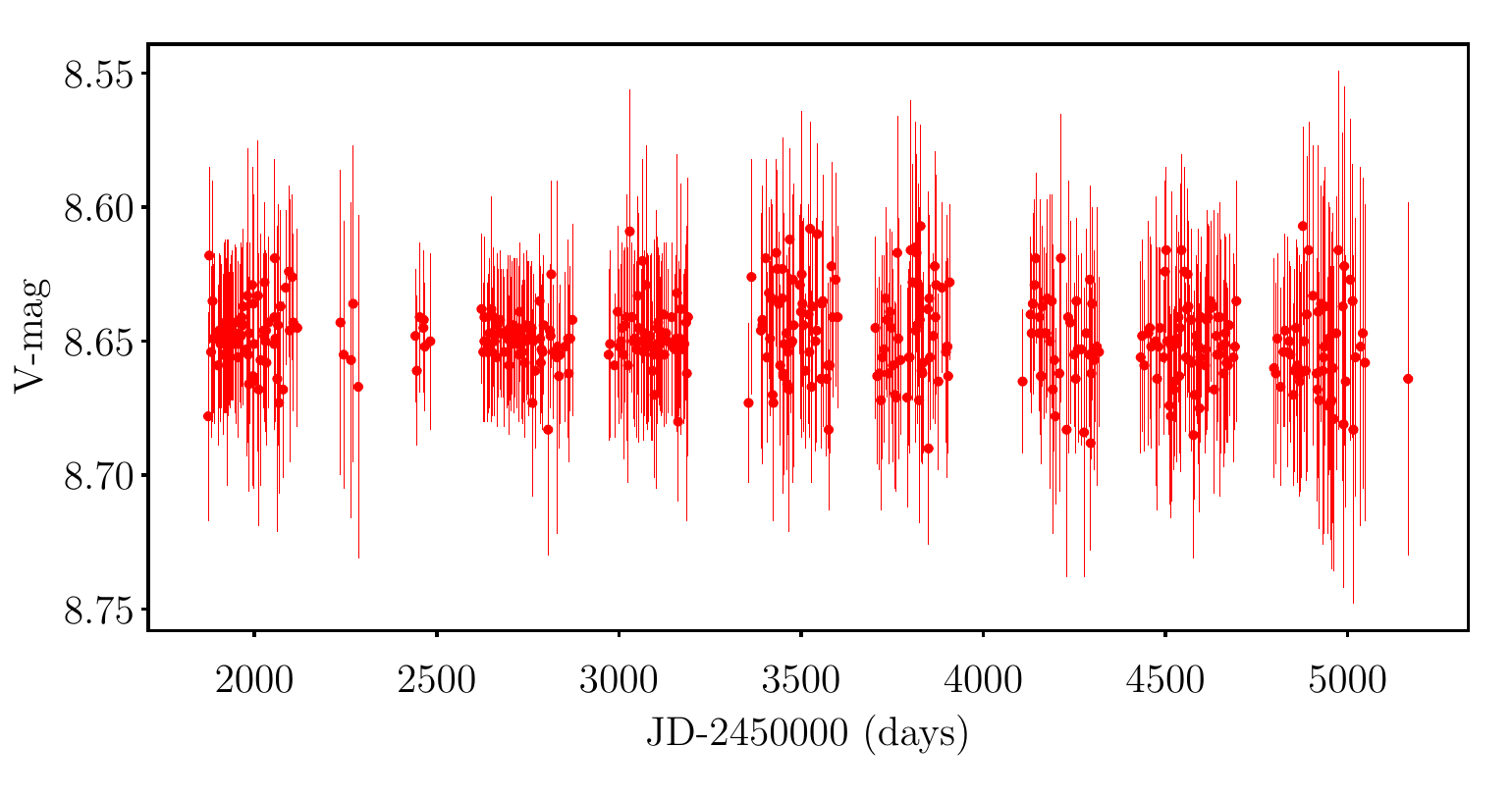}
  \label{fig:subim25}
\end{subfigure}
\begin{subfigure}{0.5\textwidth}
  \includegraphics[width=0.9\linewidth]{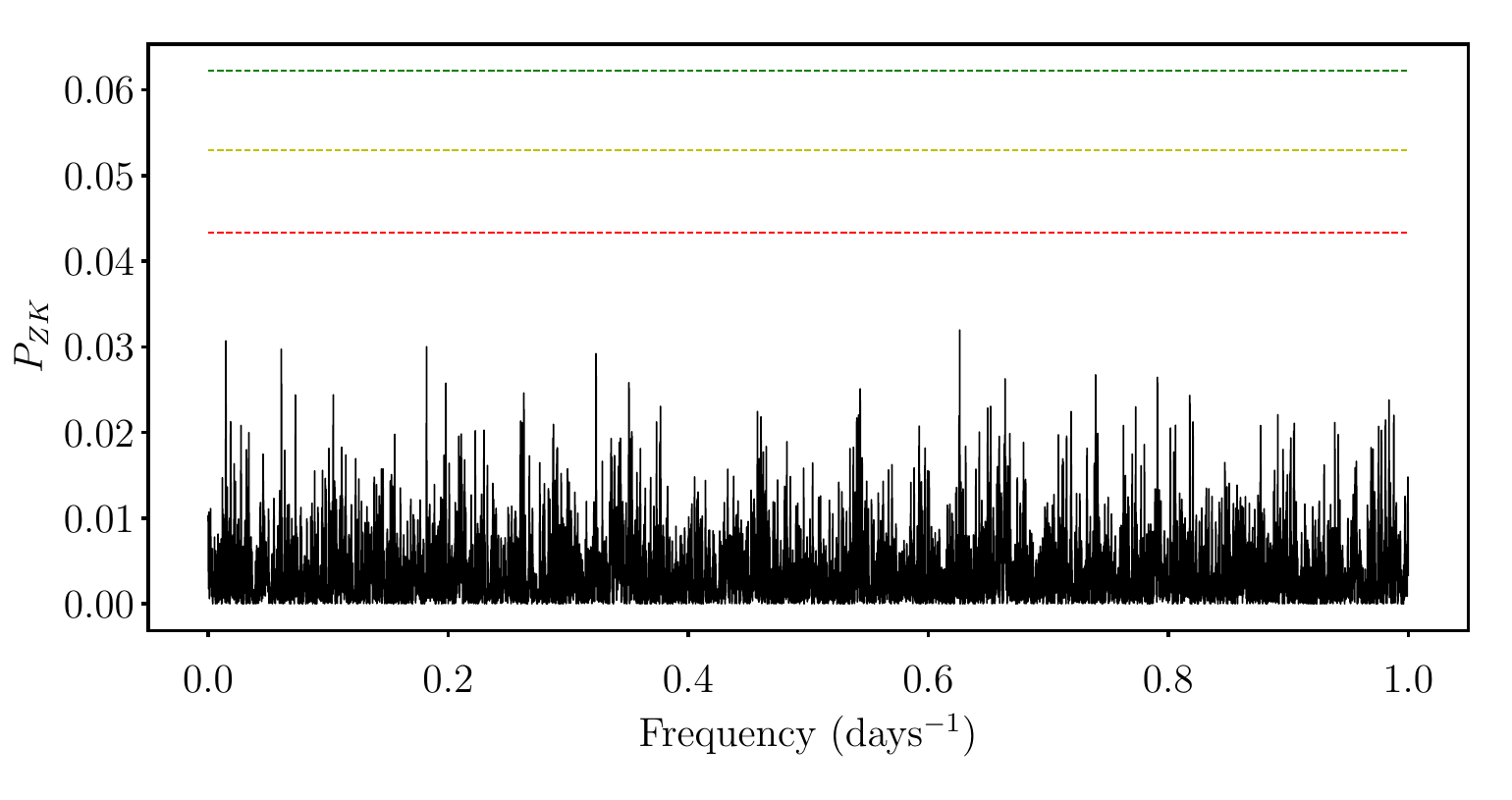}
  \label{fig:subim26}
\end{subfigure}
\begin{subfigure}{0.5\textwidth}
  \includegraphics[width=0.9\linewidth]{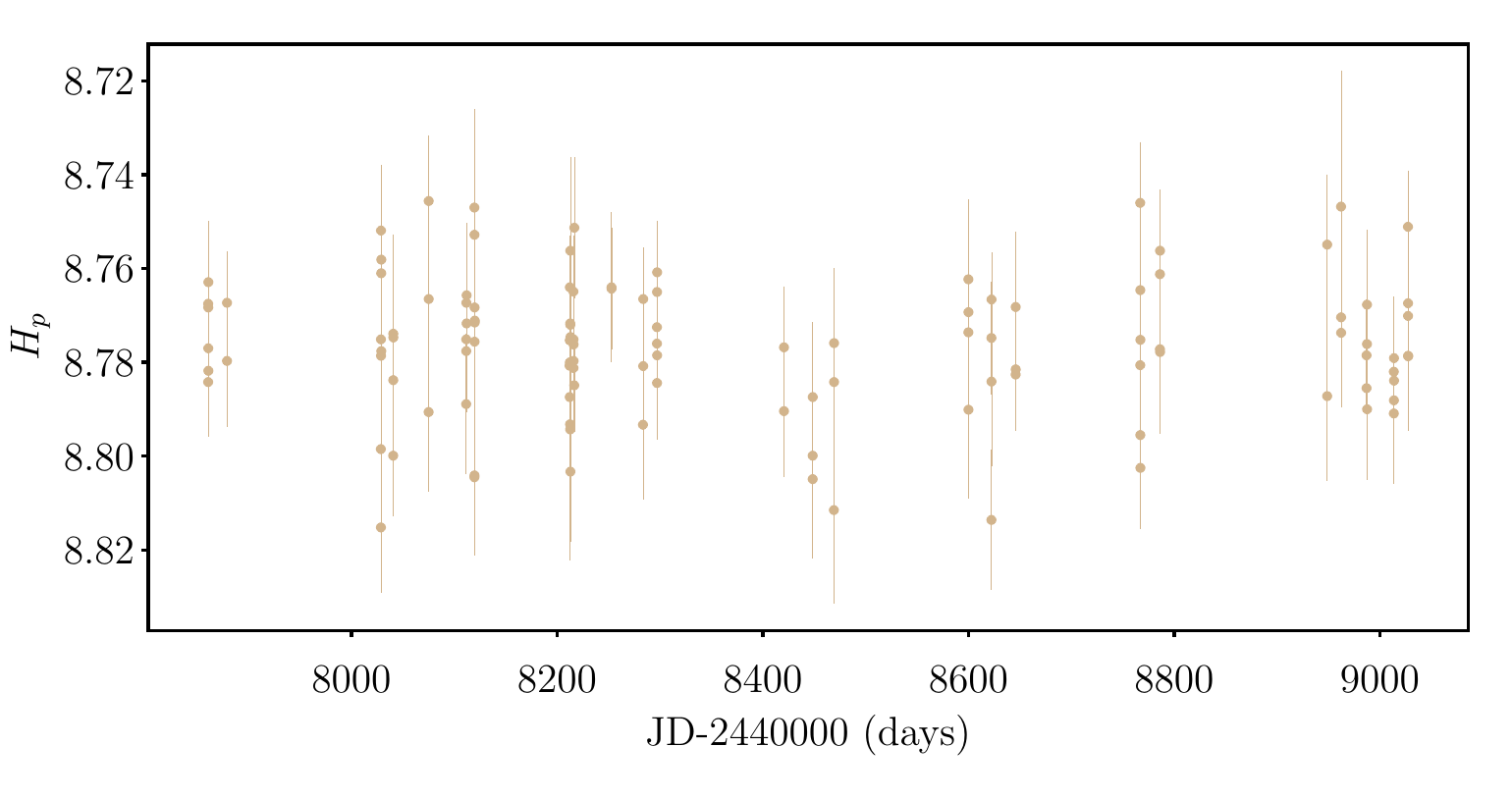}
  \label{fig:subim23}
\end{subfigure}
\begin{subfigure}{0.5\textwidth}
  \includegraphics[width=0.9\linewidth]{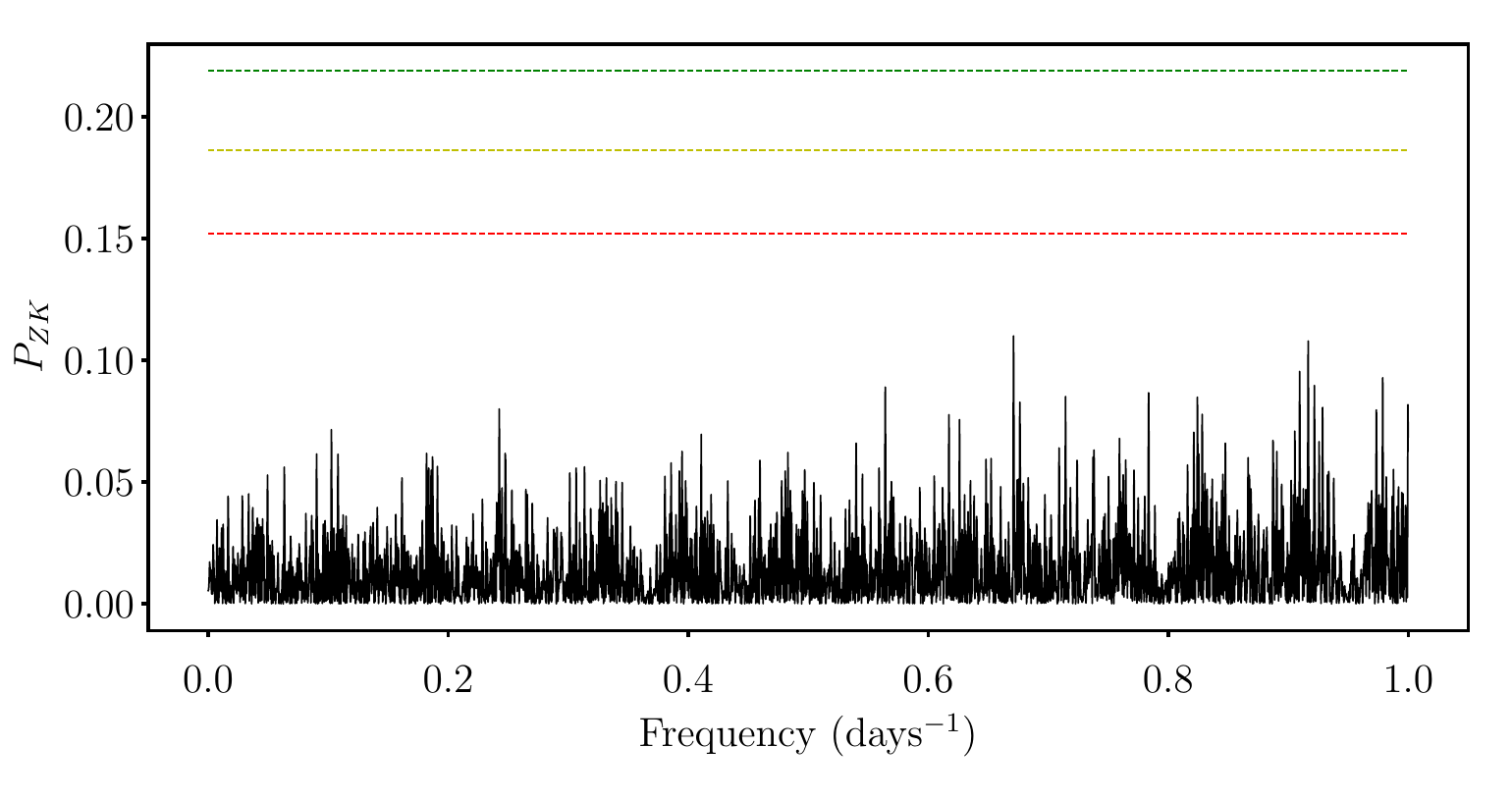}
  \label{fig:subim27}
\end{subfigure}
\begin{subfigure}{0.5\textwidth}
  \includegraphics[width=0.9\linewidth]{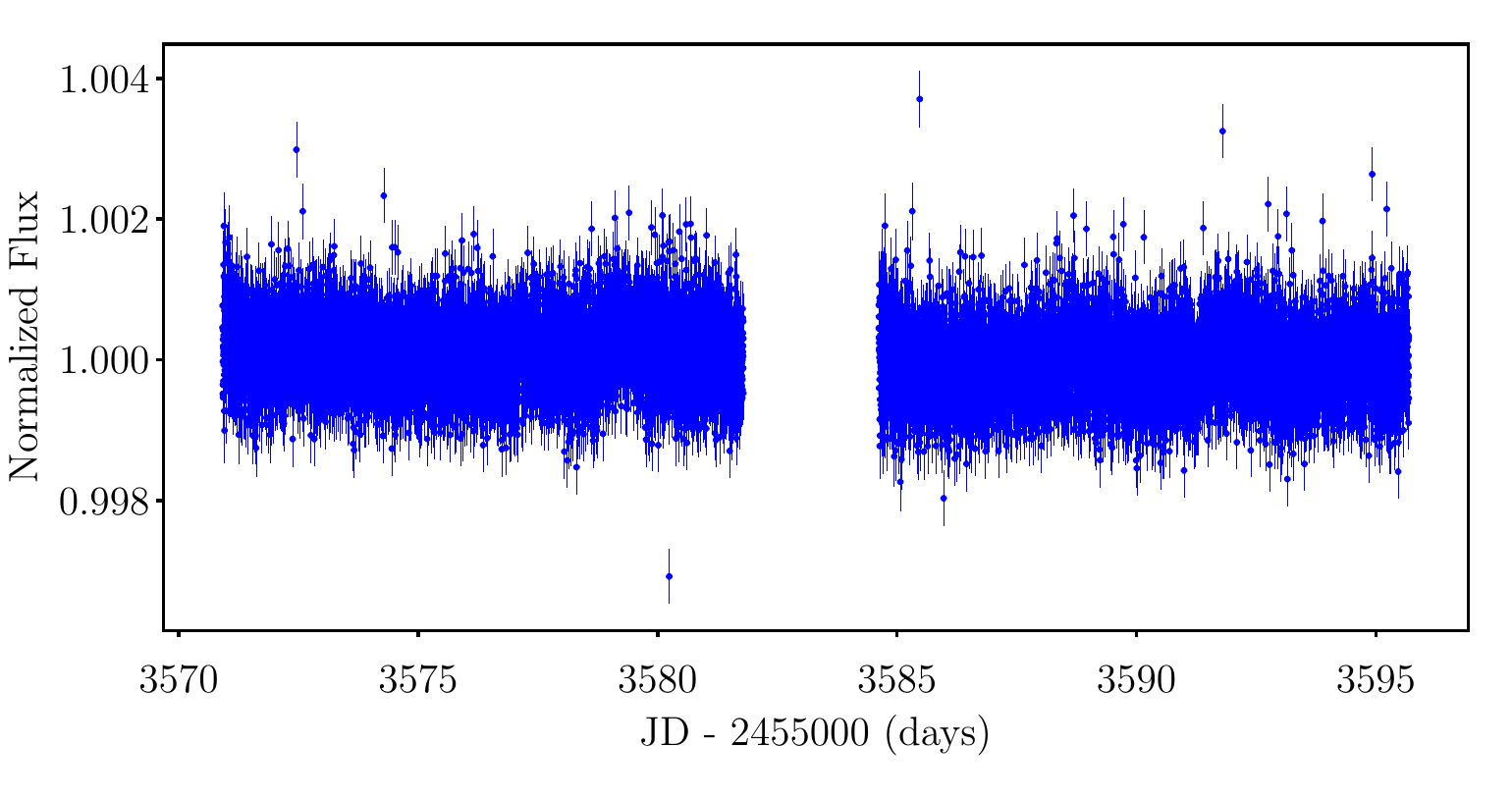}
  \label{fig:subim28}
\end{subfigure}
\begin{subfigure}{0.5\textwidth}
  \includegraphics[width=0.9\linewidth]{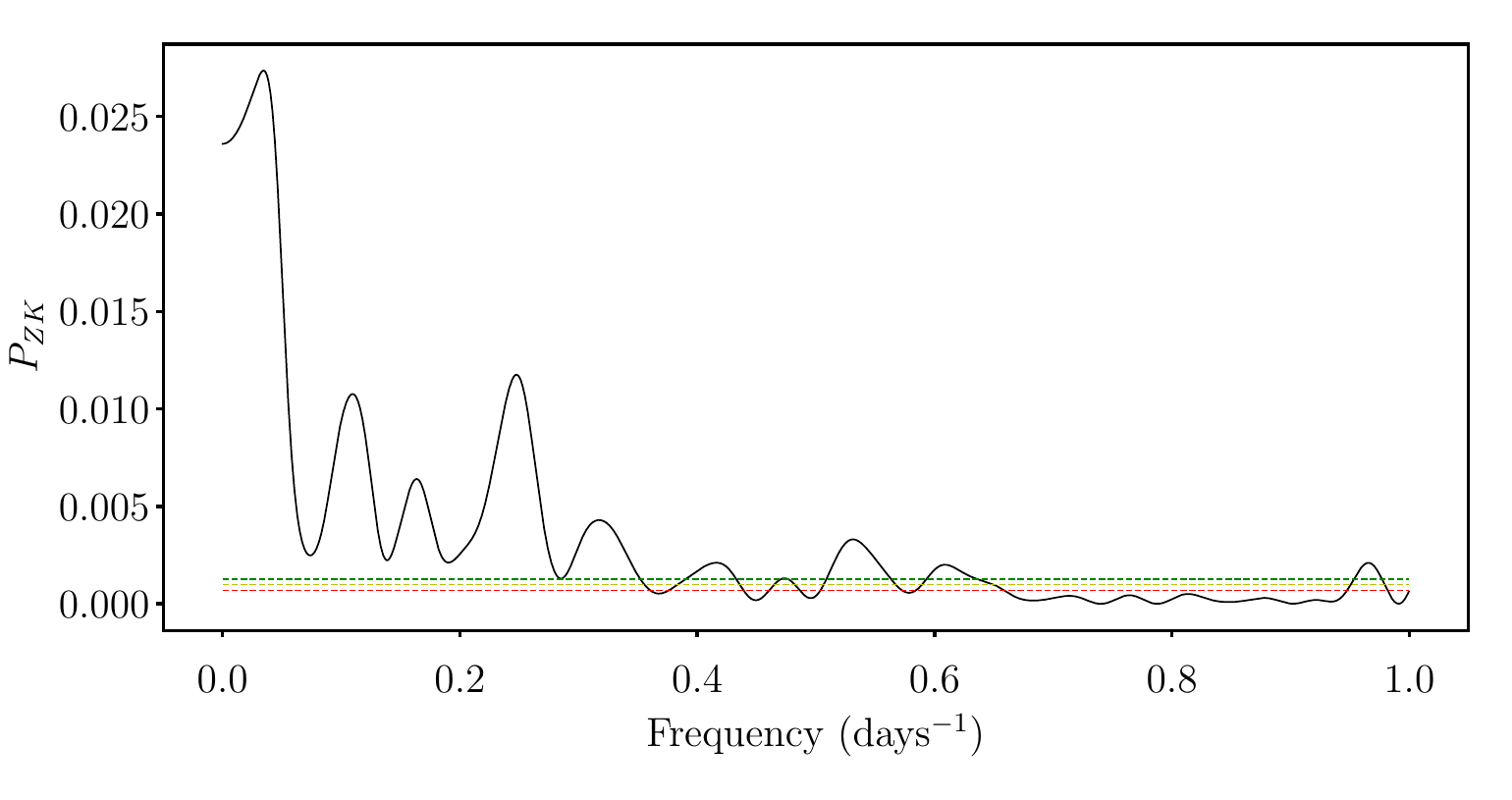}
  \label{fig:subim29}
\end{subfigure}
\begin{subfigure}{0.5\textwidth}
  \includegraphics[width=0.9\linewidth]{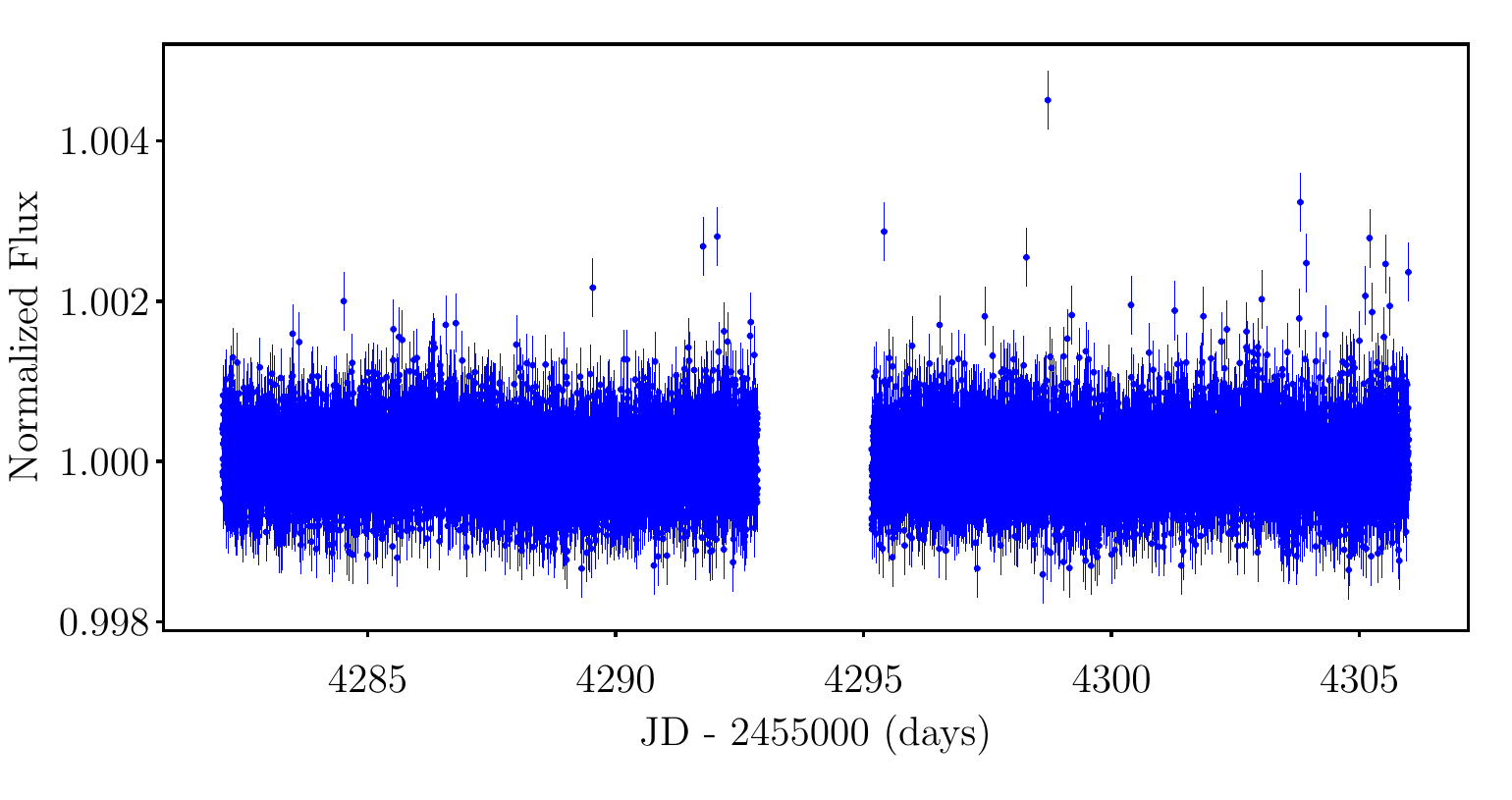}
  \label{fig:subim30}
\end{subfigure}
\begin{subfigure}{0.5\textwidth}
  \includegraphics[width=0.9\linewidth]{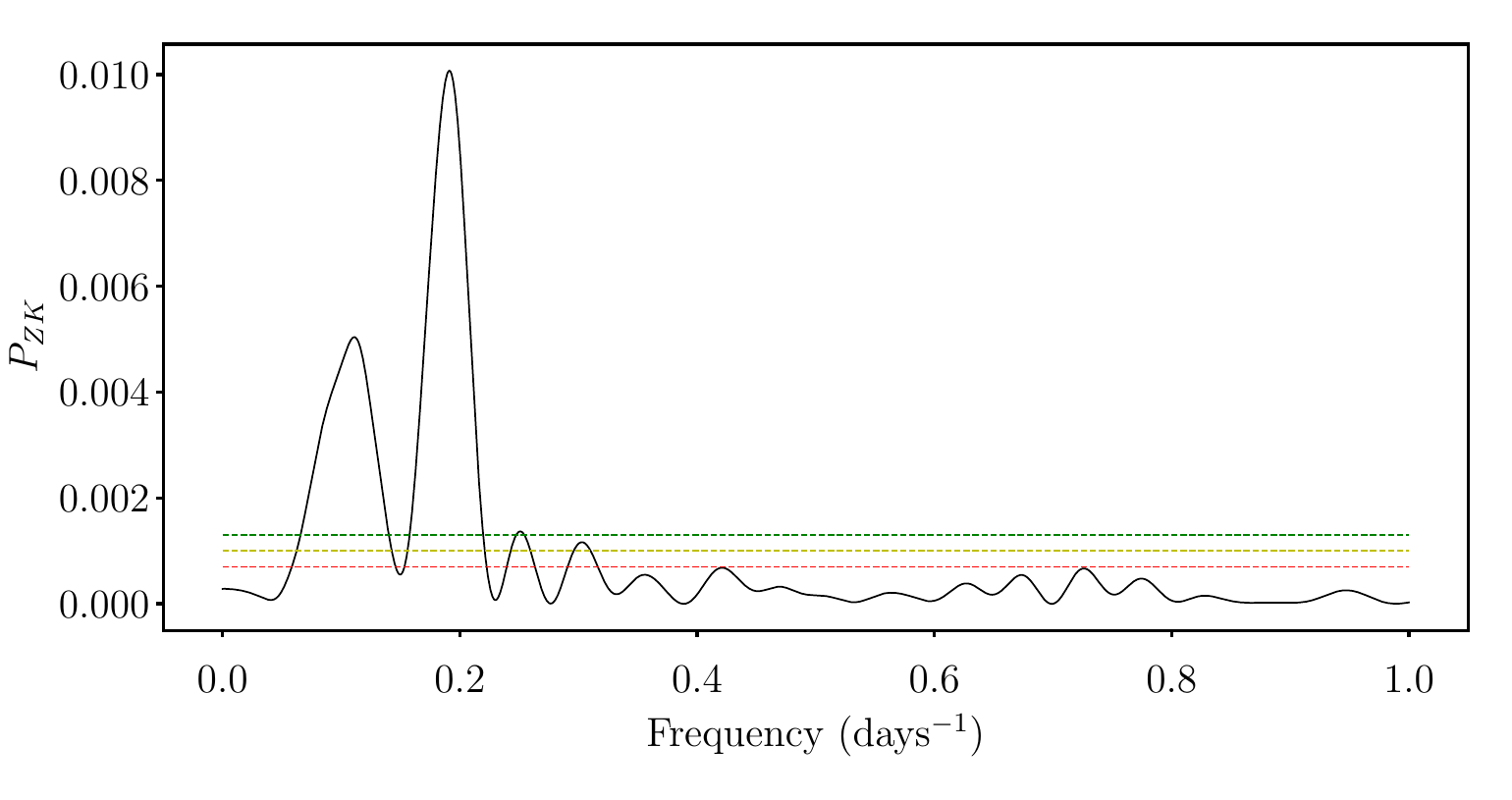}
  \label{fig:subim31}
\end{subfigure}

\caption{Photometry analysis for HD\,105779. \textit{Row 1, Left:} input ASAS V Band photometric magnitudes \textit{Row 1, Right:} GLS periodogram of ASAS data. The periodogram has a peak at $\sim1.59$\,d. \textit{Row 2, Left:} Input Hipparcos magnitudes. \textit{Row 2, Right:} GLS periodogram of Hipparcos data shows a non significant peak at $\sim1.49$\,d. \textit{Row 3, Left:} Input TESS data from sector 10. \textit{Row 3, Right:} GLS periodogram of TESS flux values shows a peak at $\sim29.13$\,d. \textit{Row 4, Left:} TESS data from sector 36. \textit{Row 4:} GLS periodogram of TESS flux values shows a peak at $\sim5.23$\,d.
}
\label{fig:a2}
\end{figure}

\onecolumn

\newpage

\subsection{Posterior samples}
\begin{figure*}[!ht]
 \centering
 \includegraphics[width = 1\linewidth,height = 16cm]{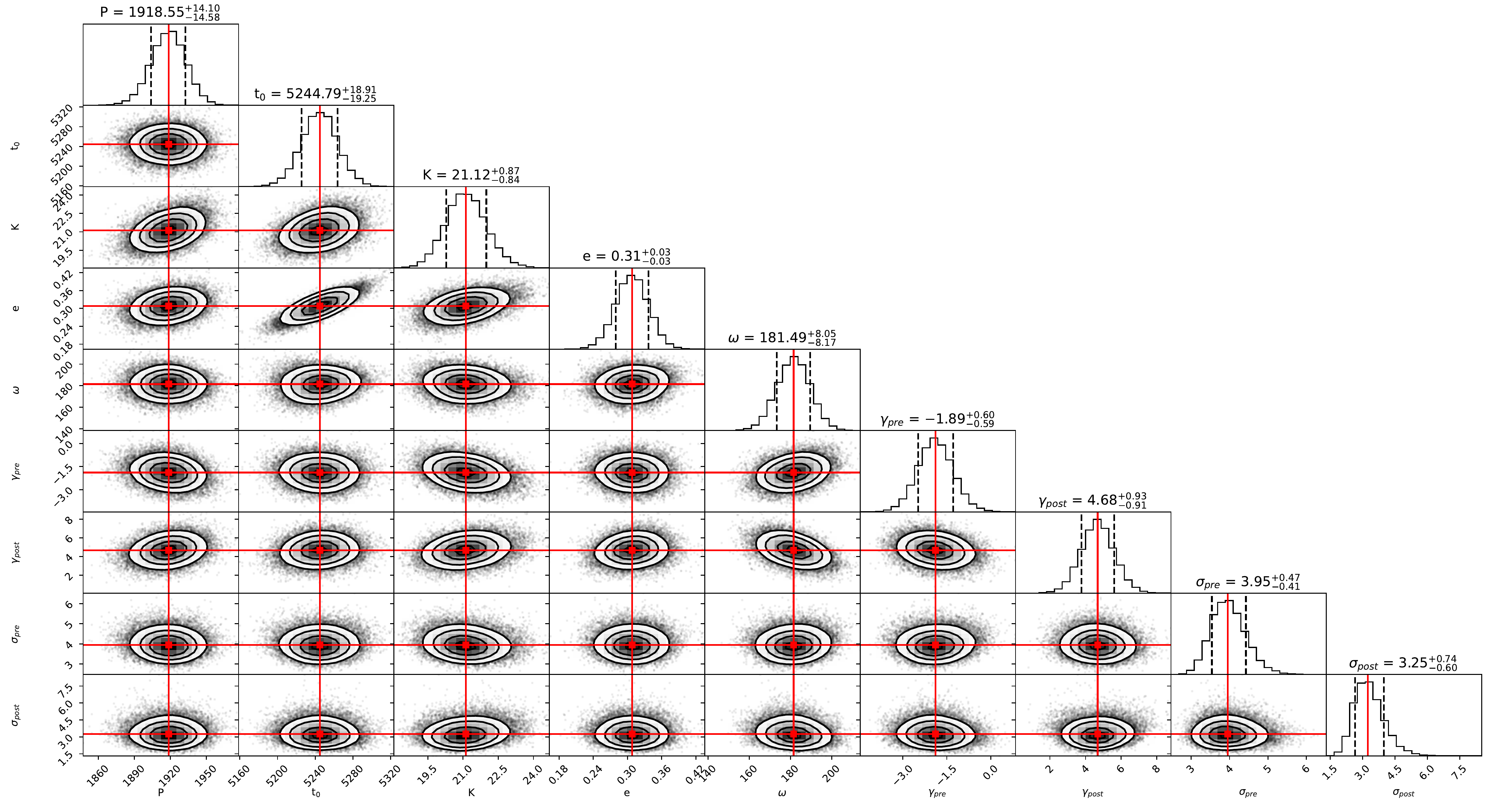}
 \caption{Posterior distributions from nested sampling for the parameters of HD\,103891\,b. Vertical lines (red) represent the median and dashed lines on histograms show the $1\sigma$ limit on parameter values.}
 \label{fig:a3}
\end{figure*}

\begin{figure*}[!ht]
 \centering
 \includegraphics[width = 1\linewidth,height = 16cm]{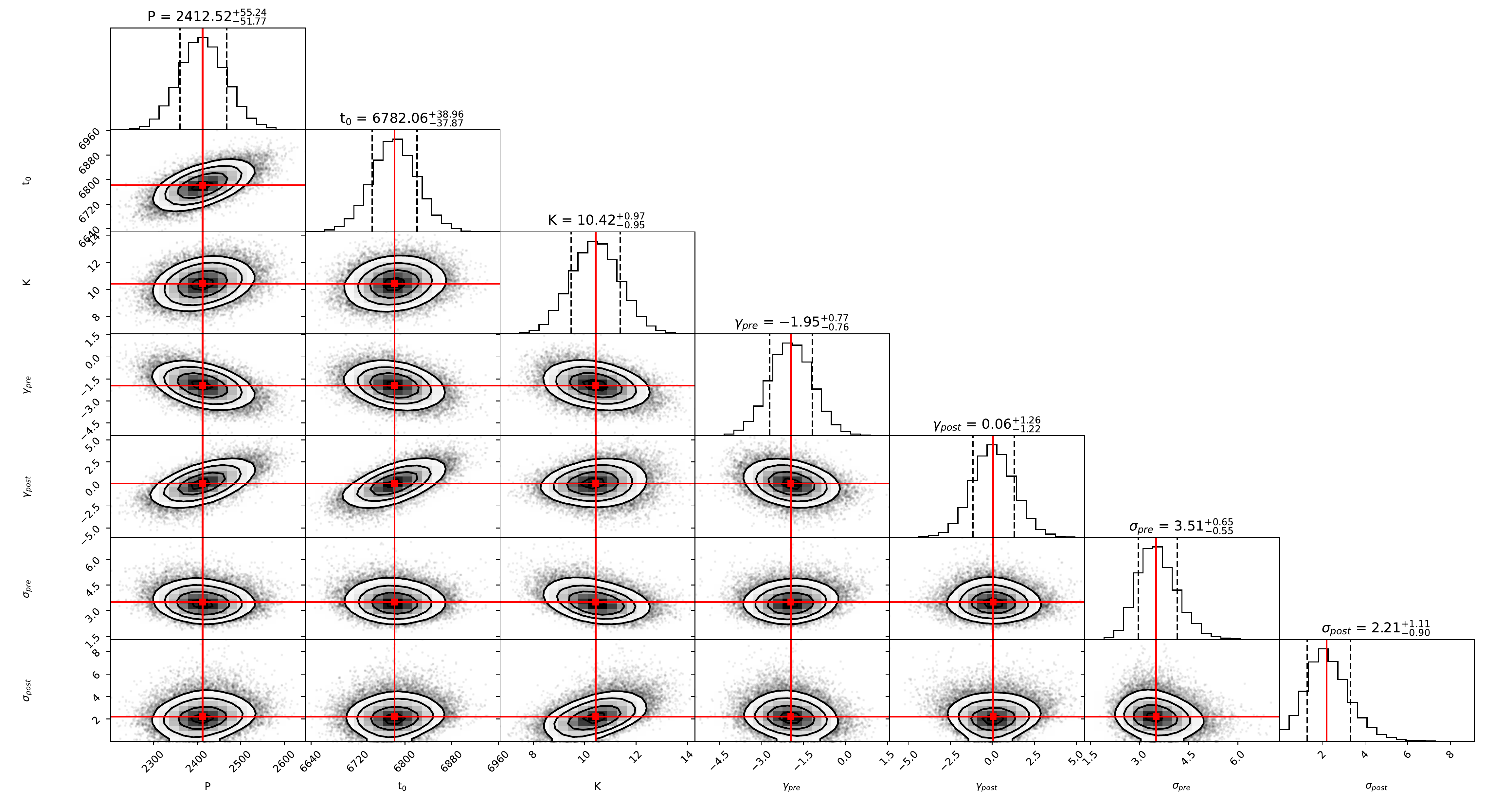}
 \caption{Posterior distribution of parameters generated using nested sampling for HD\,105779\,b.}
 \label{fig:a4}
\end{figure*}
\end{appendix}
\end{document}